\documentclass[preprint]{aastex}
\usepackage{emulateapj5,apjfonts}

\begin{document}

\def\plotthree#1#2#3{\centering \leavevmode
\epsfxsize=0.30\columnwidth \epsfbox{#1} \hfil
\epsfxsize=0.30\columnwidth \epsfbox{#2} \hfil
\epsfxsize=0.30\columnwidth \epsfbox{#3}}

\title{Extraplanar Emission-Line Gas in Edge-On Spiral Galaxies. \\ 
II. Optical Spectroscopy}

\author{Scott T. Miller\altaffilmark{1,2} and Sylvain Veilleux\altaffilmark{1,3,4}}

\affil{Department of Astronomy, University of Maryland, College Park,
MD 20742; \\ stm, veilleux@astro.umd.edu}

\altaffiltext{1}{Visiting Astronomer, Kitt Peak National Observatory
and Cerro Tololo Inter-American Observatory, National Optical Astronomy
Observatory, which is operated by the Association of Universities
for Research in Astronomy, Inc. (AURA) under cooperative agreement
with the National Science Foundation}

\altaffiltext{2}{Current address: Department of Astronomy, Pennsylvania State 
University, 525 Davey Lab., University Park, PA 16802; stm@astro.psu.edu}

\altaffiltext{3}{Current address: 320-47 Downs Lab., Caltech, Pasadena, 
CA 91125 and Observatories of the Carnegie Institution of Washington, 
813 Santa Barbara Street, Pasadena, CA 91101; veilleux@ulirg.caltech.edu}

\altaffiltext{4}{Cottrell Scholar of the Research Corporation}

\begin{abstract}
The results from deep long-slit spectroscopy of nine edge-on spiral
galaxies with known extraplanar line emission are reported.  Emission
from H$\alpha$, [N~II]$\lambda\lambda$6548, 6583, and
[S~II]$\lambda\lambda$6716, 6731 is detected out to heights of a few
kpc in all of these galaxies.  Several other fainter diagnostic lines
such as [O~I] $\lambda$6300, [O~III] $\lambda\lambda$4959, 5007, and
He~I $\lambda$5876 are also detected over a smaller scale.  The
relative strengths, centroids and widths of the various emission lines
provide constraints on the electron density, temperature, reddening,
source(s) of ionization, and kinematics of the extraplanar gas. In all
but one galaxy, photoionization by massive OB stars alone has
difficulties explaining all of the line ratios in the extraplanar
gas. Hybrid models that combine photoionization by OB stars and
another source of ionization such as photoionization by turbulent
mixing layers or shocks provide a better fit to the data. The (upper
limits on the) velocity gradients measured in these galaxies are
consistent with the predictions of the galactic fountain model to
within the accuracy of the measurements.
\end{abstract}

\keywords{diffuse radiation -- galaxies: halos -- galaxies: ISM --
galaxies: spiral -- galaxies: structure}

\section{Introduction}

Emission-line diagnostics have been used successfully to determine the
hardness of the ionizing spectrum in Galactic and extragalactic H~II
regions (e.g., Stasinska 1982; Evans \& Dopita 1985; McCall, Rybski,
\& Shields 1985; Dopita et al. 2000 and references therein) and in the
nuclei of galaxies (e.g., Baldwin, Phillips, \& Terlevich 1981;
Veilleux \& Osterbrock 1987; Osterbrock, Tran, \& Veilleux 1992;
Veilleux 2002 and references therein), but only over the last decade
has it been possible to measure the emission line ratios in the faint,
extraplanar diffuse ionized gas (eDIG) of external galaxies (e.g.,
reviews by Dettmar 1992 and Dahlem 1997).  Observations of the diffuse
ionized gas in our own Galaxy show line ratios which are difficult to
explain with pure stellar photoionization models without extra heating
(e.g., Reynolds 1985a, 1985b; Reynolds \& Tufte 1995; Mathis 2000).  A
similar situation appears to apply to external galaxies. The [N~II]
$\lambda$6583/H$\alpha$ and [S~II] $\lambda$6716, 6731/H$\alpha$ line
ratios measured in a few galaxies generaly become stronger with
increasing heights, often reaching values considerably higher than
typical values observed in H~II regions (e.g., Rand, Kulkarni, \&
Hester 1990; Keppel et al. 1991; Dettmar \& Schultz 1992; Veilleux,
Cecil, \& Bland-Hawthorn 1995; Ferguson, Wyse, \& Freeman 1996; Golla,
Dettmar, \& Domg\"orgen, 1996; Domg\"orgen \& Dettmar 1997; Rand 1998;
Otte \& Dettmar 1999; T\"ullman \& Dettmar 2000; T\"ullman et
al. 2000; Miller \& Veilleux 2003a, hereafter Paper I). The vertical
[N~II]/H$\alpha$ and [S~II]/H$\alpha$ gradients detected in these
galaxies may be due to hardening of the OB-star radiation as it passes
through the dusty and neutral medium of the galaxy, or to the
existence of other sources of heating or ionization which is becoming
increasingly important above the galactic plane. Possible sources of
extra ionization and heating include shocks, photoionization by
cooling hot gas, ``turbulent mixing layers'' (TML; Slavin, Shull, \&
Begelman 1993) or supernova remnants (Slavin, McKee, \& Hollenbach
2000), cosmic ray heating (e.g., Lerche \& Schlickeiser 1982;
Hartquist \& Morfill 1986; Parker 1992), and magnetic reconnection
(e.g., Raymond 1992).

The measurements of additional line ratios can shed some light on the
importance of secondary ionization sources. One particularly important
line ratio is [O~III] $\lambda$5007/H$\alpha$, a good indicator of
high energy processes.  However, [O~III] is challengingly faint and
has therefore been measured in only a few galaxies (e.g., Rand 1998;
T\"ullman \& Dettmar 2000; T\"ullman et al. 2000; Collins \& Rand
2001).  This is also the case for He I $\lambda$5876/H$\alpha$, a
sensitive indicator of the hardness of the ionizing
radiation. Interestingly, the value of He~I/H$\alpha$ in NGC~891, a
galaxy which in many ways is very similar to our own, appears
significantly larger than the Galactic value (0.034 {\em versus} 0.012
$\pm$ 0.006; Reynolds \& Tufte 1995; Rand 1997), while the value
observed in NGC~3044 is even larger ($\sim$ 0.07; T\"ullman \& Dettmar
2000). [O~II] $\lambda$3727/H$\beta$ has recently been shown to be a
useful diagnostic of extra heating in the diffuse ionized gas (Mathis
2000; Otte et al. 2001; Otte, Gallagher, \& Reynolds 2002), but
H$\beta$ emission is generally very faint outside of H~II regions and
[O~II] $\lambda$3727/H$\alpha$ is highly sensitive to reddening
corrections and flux calibration errors.

There is a need to expand the set of high-quality spectroscopic
observations of the eDIG to a larger number of galaxies.  This paper
describes an attempt to remedy this situation.  The results from a
spectroscopic survey of nine edge-on galaxies with known eDIG are
reported.  Due to scheduling constraints, the imaging observations
reported in Paper I were not reduced and analyzed in time for our
scheduled spectroscopic observations, so the spectroscopic sample was
selected independently of the imaging sample. The only exception is
NGC~2820, where the H$\alpha$ image obtained with the TTF (see Paper
I) was used to determine the optimum slit position [note that NGC~4013
is also in the imaging sample of Paper I, but the position of the slit
for this object is based on earlier observations by Rand (1996;
hereafter R96)].  The nine galaxies in the spectroscopic sample were
selected based on the published reports of extraplanar emission by
Pildis, Bregman, \& Schombert (1994b, hereafter PBS) and R96.  The
methods used to acquire and reduce these data are discussed in \S
2. Great care is taken to reach a limiting surface brightness of order
a few 10$^{-18}$ erg s$^{-1}$ cm$^{-2}$ arcsec$^{-2}$. The results
from the spectroscopic analysis are given in \S 3.  New constraints on
the physical conditions in the eDIG (e.g., temperature, density,
reddening, ionization level, kinematics) are derived using the
relative strengths and positions of the stronger emission lines that
lie within 4550 -- 7300~\AA.  In \S 4, the line ratios derived from
the long-slit spectra are compared with values measured in other
galaxies, as well as with predictions from photoionization models
(Sokolowski 1994; Bland-Hawthorn et al. 1997), turbulent mixing layer
models (Slavin et al. 1993), and shock models (Shull \& McKee 1979;
Dopita \& Sutherland 1995). This analysis allows us to determine
whether a secondary source of ionization in addition to
photoionization by hot stars is needed in the eDIG.  The main results
are summarized in \S 5.

\section{Data Acquisition and Reduction}

All of the data were taken at the KPNO 2.1-m telescope on January 22
-- 26, 1998.  A compromise had to be made between broad wavelength
coverage and good spectral resolution. Grating \#26new and filter
GG-420 were used with the F3KA CCD to provide a dispersion of 1.256
\AA~pixel$^{-1}$ and a useful spectral coverage of $\sim$ 2760 \AA\
between $\sim$ 4550 and 7300 \AA, after accounting for the known bad
columns on the redward side of the F3KA CCD.  The slit width was set
at 1$\farcs$5, yielding a spectral resolution of $\sim$ 3.7 \AA.  This
configuration allows us to resolve the H$\alpha$ + [N~II]
$\lambda\lambda$6548, 6583 complex, the [S~II] $\lambda\lambda$6717,
6731 doublet, and the He~I $\lambda$5876 line from the nearby Na~ID
sky lines, but does not cover the important [O~II] $\lambda$3727 and
[O~III] $\lambda$4363 diagnostic lines. The length of the slit in this
mode is 5$\farcm$2, extending well beyond the extent of the galaxy
disks in our sample. The CCD was binned in the spatial direction by a
factor of 2 in order to increase the signal per pixel from the diffuse
emission, resulting in a spatial scale of 1$\farcs$56 pixel$^{-1}$.
To reach our goal of achieving a flux limit on the order of a few
$\times$ 10$^{-18}$ erg s$^{-1}$ cm$^{-2}$ arcsec$^{-2}$, each galaxy
was observed for about 5 hours.  Details on the observations are
listed in Table 2. The slit was centered on the disk of the galaxy and
in most cases positioned so that it lay perpendicular to the disk,
although in a few cases it was tilted slightly to optimize the
coverage of the extraplanar emission. The position of the slit is
mentioned in Table 2 and shown in Figure~1 for each galaxy in the
sample. All galaxies were observe through an airmass of less than
$\sim$ 1.5 to avoid any significant differential atmospheric
refraction.

Bias frames were taken each night and a composite bias was made by
combining the individual frames.  The bias level was found not to be
constant across the CCD, showing a gradient along the dispersion axis.
A one-dimensional fit was applied along the dispersion axis to create
the bias frame, making sure not to introduce additional noise to the
data when subtracting off the bias.  The spectra were then trimmed and
corrected for bias and overscan using the CCDPROC package in IRAF.
Both dome and internal (quartz) flats were obtained during the night
with the purpose to use them for the flatfield and illumination
corrections.  However, better results were obtained when using the
domeflats to flatfield the data, and the data themselves to correct
for the illumination along the slit.  For this procedure, each
spectrum was binned along the dispersion axis so that each wavelength
bin was well represented and contained sufficient counts.  The
illumination variations across the slit were accurately modeled using
the IRAF ILLUMINATION routine to fit a spline function to the
background while avoiding the emission due to the galaxy.

Next the data were corrected for distortion and wavelength calibrated
using HeNeAr spectra.  Observations of the HeNeAr lamp were obtained
before and/or after each galaxy observation so that accurate
corrections could be obtained.  A fit to the sky background was
calculated and subtracted from each frame using sample rows on either
side of the galaxy, far enough from it to avoid the extraplanar
emission. The two-dimensional spectra were then flux calibrated using
HZ~44 as a calibrator and combined together.

After combining the different observations for each galaxy, the
redshifted emission lines were identified based on the systemic
optical velocity of each galaxy (as listed in de Vaucouleurs et
al. 1991).  The spatial dimension was further binned by two in order
to increase the signal per binned pixel in the weaker lines, and
one-dimensional spectra were extracted at different heights from the
disk plane.  The spatial dimension was binned even further at large
$\vert$z$\vert$ to help detect very faint extended line emission.  The
spectra for each galaxy are shown in Figure~2 as a function of
position along the slit.  In the case where a spectral line is not
detected, a representative spectrum from near the disk plane is
presented.  The line fluxes were determined using the SPLOT routine in
IRAF.  Single gaussian profiles were fitted to each line in order to
calculate the line flux and width and the level of the underlying
continuum emission.

\section{Results}

\subsection{Line Ratios}

The vertical profiles of up to nine diagnostic line ratios are shown
in Figure~3 for each galaxy.  In each figure, the vertical profile of
the H$\alpha$ emission is shown as the heavy solid line and the
continuum emission is shown as the dotted line.  Both have been scaled
arbitrarily for display purposes.  To prevent the figures from
becoming overcrowded, the error bars in all of the figures represent
1$\sigma$ uncertainties as determined from SPLOT.  In most galaxies,
H$\alpha$, [N~II] $\lambda$6583, and [S~II] $\lambda\lambda$6716, 6731
have been detected along the slit out to 1 -- 2 kpc from the center of
the disk (in most of our galaxies, the slit is perpendicular to the
disk and this represents the actual vertical extent of the gas, but in
a few cases this distance differs slightly from the vertical height;
see Fig. 1).  Fainter lines such as H$\beta$, [O~III] $\lambda$4959,
5007, and He~I $\lambda$5876 are also detected over a smaller scale in
a number of galaxies. This section of the paper discusses the overall
trends found in the sample. For a more detailed discussion of each
object, the reader should refer to the Appendix.

\subsubsection{[N~II]/H$\alpha$, [S~II]/H$\alpha$, and [N~II]/[S~II]}

The average midplane values for [N~II] $\lambda$6583/H$\alpha$ and
[S~II] $\lambda$6716/H$\alpha$ are 0.40 $\pm$ 0.20 and 0.24 $\pm$
0.14, respectively.  Seven galaxies show a general increase in
[N~II]/H$\alpha$ and [S~II]/H$\alpha$ with increasing height.  The
most dramatic gradients occur within NGC~4013 and NGC~4217, where the
[N~II]/H$\alpha$ ratios are H~II region-like in the disk ($\sim$ 0.3 -
0.4) but reach values of nearly 2.0 at $\vert$z$\vert$ $\sim$ 1 -- 2
kpc above the disk.  Some of the galaxies in the sample have [S~II]
$\lambda$6716/[N~II] $\lambda$6583 ratios which change with height;
some (such as NGC~2820 and UGC~4278) show a steady increase with
increasing $\vert$z$\vert$. Others (such as NGC~4302) show a general
decrease with increasing height.  Most interesting is NGC~4217, which
shows an increase in [S~II]/[N~II] from 0.5 to 0.7 up to $\vert z
\vert$ $\sim$1~kpc, and then the ratio falls to 0.3 at higher
$\vert$z$\vert$. Other galaxies present [S~II]/[N~II] ratios which are
consistent with being constant within the uncertainties of the
measurements.

For comparison, Rand (1998) found in NGC~891 that the [N~II]/H$\alpha$
line ratio rises from 0.35 in the plane of this galaxy, to a value
greater than unity at $\vert$z$\vert$ $\sim$ 2 -- 3 kpc.  A similar
trend was found with the [S~II]/H$\alpha$ line ratio, such that the
[S~II]/[N~II] ratio remained almost constant with a value $\sim$ 0.6.
Observations of a few other galaxies (Collins \& Rand 2001; T\"ullmann
\& Dettmar 2000) find similar trends.  The ratio of collisionally
excited lines (like [N~II] and [S~II]) to recombination lines (like
H$\alpha$) depends on the ratio of heating to recombination.  Since
S$^0$ and N$^0$ have similar ionization potentials (10.4 eV and 14.5
eV, respectively), any changes in the [S~II]/[N~II] ratio suggest
changes in the local ionization condition (S$^+$ ionizes at a slightly
lower ionization energy than N$^+$), or in the metallicity (since
nitrogen is a secondary product of nucleosynthesis while sulphur is a
primary product). These issues are discussed further in \S 4.2 and \S
4.3 below.

\subsubsection{He~I/H$\alpha$}

He~I $\lambda$5876 is detected in the disk of only 4 of the 9 sample
galaxies (the presence of the Na~ID absorption lines in at least 4 of
the galaxies makes detection of this line impossible - see Fig. 2 for
details).  Emission from H~II regions in the disks of the galaxies is
almost certainly contaminating some of these measurements. The
midplane value of He~I/H$\alpha$ for these galaxies ranges from 0.018
(NGC~4013) to 0.052 (NGC~4302), bracketing the value measured in Orion
(0.042; Kaler 1976) but larger than the value measured in the diffuse
ionized gas near the midplane of our Galaxy (0.012 $\pm$ 0.006;
Reynolds \& Tufte 1995).  The relative ionization fractions of helium
and hydrogen can be determined by the equation:
\begin{equation}
\frac{E_{5876}}{E_{H\alpha}} = 0.048 \frac{\chi_{He}}{\chi_H}(\frac{He/H}{0.1})(\frac{T}{8000~K})^{-0.14}
\end{equation}
(e.g., Brockelhurst 1971; Martin 1988; Reynolds \& Tufte 1995; Rand
1997), where $E$ is the emissivity in cm$^{-3}$ s$^{-1}$, $\chi$ is
the fraction of helium or hydrogen that is singly ionized, He/H is the
abundance of helium with respect to hydrogen by number, and $T$ is the
gas temperature.  Using the He/H abundance listed in Boesgaard \&
Steigman (1995) and $T$ = 8000 K (e.g., Reynolds 1992),
$\chi_{He}$/$\chi_H$ is found to range from 0.38 to 1.09.  If hydrogen
is assumed to be mostly ionized (a reasonable assumption given the
strength of [O~III] in the eDIG of several galaxies; see \S 4.2), then
the helium is about 40\% ionized in NGC~4013, and is almost fully
ionized in NGC~4302.  Using Table 1 of Rand (1997), Q$_{He}$/Q$_H$,
the ratio of He-ionizing ($h\nu$ $>$ 24.6~eV) to H-ionizing ($h\nu$
$>$ 13.6~eV) photons, ranges from $\sim$ 0.040 to 0.115.  These
results imply an effective temperature of the radiation field, $T_*$,
which ranges from about 36,500 to 38,500~K, and an upper limit to the
stellar mass function, $M_u$, which ranges from $\sim$ 42 to 54
$M_{\odot}$.

Extraplanar He~I emission is detected unambiguously in only one galaxy
of our sample, NGC~2820. This object has a midplane He~I/H$\alpha$
value of 0.046, and a range of values from 0.033 to 0.056 (the latter
at a height of $\vert$z$\vert$ = 1~kpc).  Corresponding values of
[N~II]/H$\alpha$ are $\sim$ 0.25 at the midplane, and range from 0.16
to $\sim$ 0.3 at $\vert$z$\vert$ = 1~kpc.  Within the uncertainties,
these values are consistent with the predictions from the O-star
photoionization models of Domg\"orgen \& Mathis (1994).  Rand (1997)
had difficulties reconciling the values of He~I/H$\alpha$ with those
of [N~II]/H$\alpha$ in NGC~891, but the extraplanar [N~II]/H$\alpha$
line ratios in this galaxy are much higher (near 1.4).  Extraplanar
He~I was not detected in any of the galaxies of our sample with
large extraplanar [N~II]/H$\alpha$ ratios. 

\subsubsection{Reddening from H$\alpha$/H$\beta$}

The H$\alpha$/H$\beta$ line ratio is shown in the bottom left panel of
Figure~3 for each galaxy (when detected).  No correction was made for
possible underlying stellar absorption features; these features were
never evident in the data.  This line ratio provides an indication of
the amount of reddening affecting the spectra. Determining the amount of
reddening is not straightforward, however.  Typical H$\alpha$/H$\beta$
line ratios for H~II regions is $\sim$ 2.85, but if shocks are
present, this ratio could rise to $\sim$ 3.10 (e.g., Shull \& McKee
1979).  The effects of extinction and reddening also depend on the
distribution of the dust with respect to the source of emission (e.g.,
uniform screen in front of the source of emission {\em versus} uniform
dust distribution mixed with the line emitting gas). Rather than to
try to correct for these complex effects, we instead list in
Table~3 the impact on the line ratios of a foreground screen of dust
with $A_V$ = 1. Except for H$\alpha$/H$\beta$ and [O~III]/H$\alpha$,
the line ratios shown in Figure~3 are not at all sensitive to
reddening.

In general the amount of reddening is found to be larger in the disk
than in the eDIG, as one would expect if the dust is distributed near
the plan of the galaxy disk.  Out of the five galaxies in which
extraplanar H$\beta$ was detected, three clearly show this trend.  In
the other two galaxies, the H$\alpha$/H$\beta$ line ratio appears to
be relatively constant with height (NGC~2820) or does not show any
obvious monotonic trend with height (NGC~4013).

\subsubsection{Density from [S~II] $\lambda$6716/[S~II $\lambda$6731}

The [S~II] $\lambda$6716/[S~II] $\lambda$6731 ratio is shown in the
bottom middle frame of Figure~3 for each galaxy (when detected).  The
ratio of the intensities of these lines yield information on the
average electron density of the gas.  The low density limit of this
line ratio is $\sim$ 1.4, and so it is possible to make quantitative
statements regarding the density of the gas only in regions where this
line ratio is less than $\sim$ 1.4. 

For most of the galaxies, the [S~II] $\lambda$6716/[S~II]
$\lambda$6731 line ratio is consistent with the low density
limit, therefore suggesting an electron density of at most a few tens
of cm$^{-3}$. This is typical of the electron density that has been
reported for other galaxies (e.g., Rand 1998; Collins \& Rand 2001).
There are three exceptions: NGC~3628, NGC~4217, and NGC~4302.  The
results on NGC~3628 are discussed in the Appendix.  In NGC~4217, the
[S~II] $\lambda$6716/[S~II] $\lambda$6731 line ratio has a value of
$\sim$ 1.2 near the disk of the galaxy, and drops slightly to an
average value of 0.9 at a height of about 1 kpc, before climbing back
up over 1.4.  Assuming a constant temperature of $\sim$ 10$^4$ K for
the moment, this suggests that the electron density in the disk of the
galaxy is $\sim$ 200 cm$^{-3}$ and increases to a value of 900
cm$^{-3}$ at heights $\sim$ 1 kpc, before decreasing below the low
density limit.  In NGC~4302, the value of the [S~II]
$\lambda$6716/[S~II] $\lambda$6731 line ratio near the disk of the
galaxy is $\sim$ 1.0, and decreases to values near 0.8 and 0.6 at
heights of $\sim$ 0.5 kpc.  The corresponding electron densities are
$\sim$ 500 cm$^{-3}$ in the disk of the galaxy and 1000 to 2000
cm$^{-3}$ at larger heights.  It should be noted, however, that the
electron density as measured by the [S~II] ratio scales as T$^{1/2}$
(e.g., Osterbrock 1989), and therefore an increase in temperature
would be interpreted as higher electron density if constant
temperature were assumed. The vertical temperature profiles in the
eDIG of these galaxies are discussed next.

\subsubsection{Temperature from [N~II]/H$\alpha$}

Two of the best temperature gauges for the extraplanar gas are the
[N~II] $\lambda$6583/[N~II] $\lambda$5755 and
[O~III]$\lambda$5007/[O~III] $\lambda$4363 line ratios.  [N~II]
$\lambda$5755 was recently detected in the diffuse ionized gas of our
Galaxy and indicates elevated temperatures relative to those of H~II
regions (Reynolds et al. 2001).  Unfortunately the [N~II]
$\lambda$5755 line was not detected in any of the sample galaxies and
[O~III] $\lambda$4363 falls outside the wavelength range of our
observations. We therefore have no choice but to use other line ratios
for this analysis. Recent studies have suggested the use of the
[N~II]/H$\alpha$ line ratio as a temperature diagnostic (Haffner,
Reynolds, \& Tufte 1999), keeping in mind that the derived temperature
is an average value over the line-of-sight column density.  Given that
hydrogen and nitrogen have similar first ionization potentials and a
weak charge-exchange reaction, and assuming that hardly any N is
doubly ionized, one has N$^+$/N$^0$ $\approx$ H$^+$/H$^0$. Under this
assumption and using a Galactic gas-phase abundance of (N/H) = 7.5
$\times$ 10$^{-5}$ at all heights (Meyer, Cardelli, \& Sofia 1997),
the relationship between the [N~II]/H$\alpha$ line ratio and electron
temperature is given by
\begin{equation}
\frac{[N~II]}{H\alpha} = 12.2~T^{0.426}_4e^{-2.18/T_4}
\end{equation}
where $T_4$ is the electron temperature in units of 10$^4$ K (Collins
\& Rand 2001).  The detection of significant [O~III] emission in the
eDIG of four galaxies of our sample ([O~III]/H$\alpha$ $\ga$ 0.25;
UGC~4278, NGC~2820, NGC~4302 and NGC~5777) is inconsistent with the
[N~II]/H$\alpha$ temperature model (eqn. 2) because this model assumes
that N$^{++}$ is not present in the eDIG, yet the ionization potential
of N$^+$ (29.6~eV) is close to that of O$^+$ (35.1~eV).  Temperature
profiles for the remaining five galaxies in the sample have been
calculated using eqn. (2) and the measured [N~II]/H$\alpha$ line
ratios; the results are shown in the bottom, left panel of Figure~4
(the other panels will be discussed in \S 4.2). Three of these
galaxies (NGC~3628, NGC~4013 and NGC~4217) show an increase in
temperature with vertical distance from the disk.  The change in
temperature with height varies from galaxy to galaxy, with some
objects showing no change with height within the uncertainties (e.g.,
UGC~2092 and UGC~3326), while others show a dramatic increase (e.g.,
NGC~4217: from 6000 K to almost 10,000 K).  Haffner et al. (1999)
detect an increase in electron temperature for our Galaxy from 7000~K
at $\vert$z$\vert$ = 0.75~kpc to over 10,000~K at $\vert$z$\vert$ =
1.75~kpc.  Collins \& Rand (2001) find similar temperature gradients
in their studies of edge-on galaxies.

\subsection{Kinematics}

The kinematics of the extraplanar gas can also provide insights into
its nature and origin.  To our knowledge, only two galaxies (NGC~891
and NGC~5775) have so far been examined with sufficient care to
address this important issue. In both cases, the velocity of the
extraplanar material is seen to approach the systemic velocity of the
galaxy (Pildis, Bregman, \& Schommer 1994a; Rand 1997, 2000; T\"ullman
et al. 2000). This vertical velocity gradient can be explained by a
combination of radial movement of the gas (following the pressure
gradient of the halo) and declining rotation speed (conserving angular
momentum), as predicted by the galactic fountain model (e.g., Bregman
1980; Houck \& Bregman 1990)

Although our spectroscopic setup was chosen to optimize wavelength
coverage at the expense of spectral resolution to allow us to carry
out a detailed line ratio analysis of our objects, an attempt is made
to constrain the kinematics of the eDIG in the galaxies of our sample.
Figure~5 displays the velocity profiles for H$\alpha$, [N~II]
$\lambda$6583, and [S~II] $\lambda$6716, the strongest lines in our
galaxies.  These velocity profiles are not expected to differ
significantly from each other, so comparisons between the three panels
can serve to estimate the uncertainties of the velocity
measurements. Galactic rotation causes a velocity offset from systemic
in the cases where the slit does not pass through the nucleus.  Figure
5 shows that most galaxies do not show significant vertical gradients
within the accuracy of the measurements ($\sim$ 30 -- 50 km s$^{-1}$
depending on the galaxy).  However, there appears to be a few
exceptions. A significant dip in velocity ($\delta v \approx$ 100 km
s$^{-1}$) is observed at $z \approx -0.3$ kpc in NGC~3628, coincident
with the prominent dust lane in this object. Dust obscuration is
severely limiting the use of the optical emission lines as kinematic
probes in this region. Possibly significant gradients are also visible in
NGC~2820 and NGC~4013 (and perhaps in NGC~4302 but at a lower
significance level). In NGC~2820, the velocities reach a maximum of
$\sim$ 150 km s$^{-1}$ around $z = +0.4$ kpc and then show a monotonic
decrease toward systemic velocity at large heights, reaching values of
$\sim$ 80 km s$^{-1}$ at $z \approx~^{+1.5}_{-1.0}$ kpc.  A slightly
less significant gradient of $\sim$ 50 km s$^{-1}$ appears to be
present on both sides of the disk of NGC~4013 out to $\vert z \vert$
$\approx$ 1 kpc.  The gradients observed in both of these galaxies can
be explained if the rotational velocity of line-emitting material is
lower above and below the galaxy disk, as expected in the galaxy
fountain model. The amplitudes of the detected gradients are however
larger than the predictions from the model of Bregman (1980).  The
ballistic model of Collins, Benjamin, \& Rand (2002) has more success
explaining the gradients in NGC~2820 and NGC~4013. The lack of obvious
gradients in the other galaxies is not inconsistent with the
predictions of the galactic fountain model of Bregman (1980), given
the relatively large uncertainties in the measurements.

The widths of the emission lines provides another constraint on the
gas kinematics. The lower right panels of Figure 5 show the line
widths of H$\alpha$, selected because it is generally the strongest
emission line in our galaxies.  Typical values lie between $\sim$ 100
and 200 km s$^{-1}$.  There is little or no evidence for significant
line width gradients in the majority of the galaxies in our sample.
The only galaxies where gradients may be present are NGC~3628
(coincident with the dust lane), UGC~3326 (the line widths vary from
$\sim$ 275 km s$^{-1}$ in the midplane to $\sim$ 150 km s$^{-1}$ at
heights near $\vert$z$\vert$ = 1~kpc), NGC~4013 (a slight positive
gradient of $\sim$ 50 km s$^{-1}$ over $\pm$ 0.5 kpc on both sides of
the disk may be present), and NGC~2820 (the line widths in this object
are constant over most of the slit, but show a significant decrease
from 170 km s$^{-1}$ at $\vert$z$\vert$ $\sim$ 1~kpc to $\sim$ 80 km
s$^{-1}$ at $\vert$z$\vert$ $\sim$ 1.7~kpc in the southern halo of the
galaxy).  Numerical broadening arising from Poisson noise, whereby
noisier signals tend to pull in more emission in the wings than at the
peak, may affect (increase) the line widths at large $\vert z
\vert$. Positive vertical line width gradients may also be due to
increasingly turbulent motions at large heights (e.g., turbulent
mixing layer; \S 4.3.1), or due to the fact that extinction by the
disk is less significant at large heights (\S 3.1.3) and therefore a
longer column of material with a broader range of velocities is being
sampled. Our data do not allow us to distinguish between these various
possibilities.

\section{Discussion}

Several models have been proposed to explain the line ratios detected
in the extraplanar material of disk galaxies. In this section we
discuss each of these models and compare their predictions with our
data.

\subsection{Photoionization by OB Stars}

OB stars are almost certainly contributing to the ionization of the
eDIG.  They are by far the main source of Lyman continuum flux
produced in the disk (e.g., Reynolds 1984), but their position near
the disk midplane makes them highly vulnerable to absorption by the
ISM and dust.  Photoionization models (e.g., Mathis 1986; Domg\"orgen
\& Mathis 1994; Sokolowski 1994; Bland-Hawthorn et al. 1997; Mathis
2000) have had some success explaining the increase in the
[N~II]/H$\alpha$ line ratio observed in the eDIG.  This increase is
attributed to a decrease with height of the ionization parameter
($U$), a measure of the ratio of the ionizing photon number density
($\Phi$) to the electron density ($n_e$).  Under the assumption of
ionization equilibrium, $\Phi$ $\propto$ $n^2_e$ at all heights.
Therefore, $U$ $\propto$ $\Phi/n_e$ $\propto$ $n_e$, so the ionization
parameter should fall off exponentially with height.  As $U$
decreases, lower ionization species are favored, leading to an
increase in the [N~II]/H$\alpha$ and [S~II]/H$\alpha$ ratios and a
decrease in the [O~III]/H$\alpha$ ratio (neglecting the effects of
reddening).

However, some problems arise with the photoionization models.  First,
pure stellar continua models have difficulty reproducing
[N~II]/H$\alpha$ and [S~II]/H$\alpha$ ratios greater than unity, as
detected in some of the galaxies of our sample and in other studies
(e.g., Dettmar \& Schultz 1992; Veilleux et al. 1995; Rand 1998;
Collins \& Rand 2001; Paper I).  Photoionization models that take into
account the multi-phase nature of the ISM, the possible depletion of
certain gas-phase abundance of metals onto dust grains, and the
absorption and hardening of the stellar radiation field as it
propagates through the dust and H~I gas in the disk (Sokolowski 1994;
Bland-Hawthorn et al. 1997) are more successful at producing elevated
[N~II]/H$\alpha$ ratios of $\sim$ 1.5.  Unfortunately, these models
have difficulties explaining the observed behavior of [O~I] and
[O~III] relative to H$\alpha$.  Collins \& Rand (2001) detect [O~III]
in three out of four of their galaxies, and observe in each one a
general {\em increase} in the [O~III]/H$\alpha$ line ratio with
increasing height.  Out of the five galaxies in which we detect
[O~III], three of them (NGC~2820, NGC~4302, and NGC~5777) show the
same positive trend with height.  A sharp increase in the
[O~I]/H$\alpha$ line, which is difficult to explain using
photoionization models, is also observed in two galaxies of our sample
(NGC~2820 and NGC~4217; see also Collins \& Rand 2001; Rand 1998).  A
secondary source of heating and/or ionization appears to be needed to
explain these observations.

\subsection{Secondary Source of Nonionizing Heating}

We first explore the possibility of a nonionizing source of heating in
the eDIG (e.g., photoelectric heating from dust grains, dissipation of
interstellar turbulence). The apparent temperature gradients found in
three galaxies in the sample (\S 3.1.5) bring support to this scenario
but do not prove it. For this, one needs to also examine the behavior
of the other line ratios.  Using the assumptions mentioned in \S 3.1.5
when deriving equation (2), Galactic gas-phase abundances of S, N, and
O, and the fact that the ionization fractions of oxygen and hydrogen
are coupled through a charge exchange reaction, Collins and Rand
(2001; also Haffner et al. 1999) derive the following equations for
the line ratio intensities as a function of temperature and ionization
fraction:
\begin{equation}
\frac{[S~II]}{H\alpha} = 14.3 \left[\frac{S^+/S}{H^+/H}\right]T^{0.307}_4e^{-2.14/T_4},
\end{equation}

\begin{equation}
\frac{[S~II]}{[N~II]} = 1.2 \left[\frac{S^+/S}{H^+/H}\right]T^{-0.119}_4e^{0.04/T_4},
\end{equation}

\begin{equation}
\frac{[O~III]}{H\alpha} = 40 \left[\frac{O^{++}/O}{H^+/H}\right]T^{0.52}_4e^{-2.87/T_4},
\end{equation}

\begin{equation}
\frac{[O~I]}{H\alpha} = 7.9 \left[\frac{1-(H^+/H)}{(H^+/H)}\right]\left[\frac{T^{1.85}_4}{1+0.605T^{1.105}_4}\right] e^{-2.284/T_4}.
\end{equation}
These equations implicitly assume that the metallicity of the eDIG is
constant with height. Note that eqn. (4) predicts only a very weak
temperature dependence for the [S~II]/[N~II] ratio. Using the
temperature gradients derived in \S 3.1.5, the line ratios can be
predicted for a number of ionization fractions and compared with the
measured quantities; this is done in Figure~4.  The variations of the
[S~II]/[N~II] line ratios in UGC~3326, NGC~3628, NGC~4013, and
NGC~4217 (the large uncertainties in the data of UGC~2092 prevent us
from making any statement for this object) cannot be simply explained
by a temperature gradient with height.  The abundance of N$^+$
relative to that of S$^+$ varies with heights, perhaps an indication
that an additional source of ionization is present in these objects.

The positive gradients in the vertical [O~III]/H$\alpha$ profiles of
NGC~2820, NGC~4302, and NGC~5777 strongly suggest the presence of an
additional source of ionization that becomes more influential at lower
densities. This possibility has also been suggested to explain the
line ratios in our Galaxy (Haffner et al. 1999; Reynolds et al. 1999)
and a few other external galaxies (Collins \& Rand 2001). The next
section considers this scenario in more detail.

\subsection{Secondary Source of Ionization}

A wide variety of secondary sources of ionization have been suggested
to account for the line ratios in the eDIG (see references in \S 1).
In this section we discuss the predictions from three of the possible
scenarios and compare them to our data. The promising possibility that
cooling supernova remnants also contribute to the ionization of the
eDIG (Slavin, McKee, \& Hollenbach 2000) is not discussed here because
the predicted line ratios are not available in this case. 

\subsubsection{\em Turbulent Mixing Layers}

Slavin et al. (1993) model the turbulent mixing layer (TML) created by
shear flows between hot and cold gas. This layer is made up of
intermediate-temperature gas which radiatively cools until the cooling
rate is balanced by the energy flux into the gas layer.  The
properties of the TML depend on two main parameters, namely the
temperature attained by the gas immediately after mixing and the
enthalpy flux per particle into the layer (which itself depends on the
velocity of the hot gas, $v_t$).  Their models span a temperature
range of 5.0 $\le$ log $T$ $\le$ 5.5 and a velocity range of 25 km
s$^{-1}$ $\le$ $v_t$ $\le$ 100 km s$^{-1}$.  In their models, Slavin
et al. assume pressure and ionization equilibrium between the hot and
cold gas phases and slow mixing within the turbulent layer, so as to
distinguish between turbulent mixing and shocks.

In general, the TML models are successful at predicting many of the
observed line ratios, such as [N~II]/H$\alpha$ and [S~II]/H$\alpha$.
The TML models also predict strong [O~III] emission (due to the high
level of ionization in the mixing layer region), an emission line
which is difficult to explain with photoionization alone (\S 4.1).

Figure~6 presents plots of the [S~II]/H$\alpha$, [O~III]/H$\alpha$ and
[O~I]/H$\alpha$ line ratios {\em versus} [N~II]/H$\alpha$ for each of
the galaxies in the sample.  In the first column, the predicted values
of the line ratios based on Sokolowski's photoionization model (using
an absorption-hardened ionizing spectrum and dust-depleted abundances
in the gas phase) are represented by a series of points connected by a
thin solid line. The open triangles in these panels represent the
predictions from Slavin et al.'s dust-depleted abundance TML
models. Best-fit hybrid models that linearly combine the line ratio
predictions from pure photoionization and pure TML models are shown as
the thick solid line.  The best fit is determined visually as the set
of points that most closely match the data points in all three plots.

Of the three lefthand panels in Figure 6, the most informative is
arguably the [O~III]/H$\alpha$ {\em {\em vs.}} [N~II]/H$\alpha$
diagram.  The three TML models are clearly separated in this
diagnostic diagram.  In the other two panels, the predictions for the
log~$T$~=~5.3 and the log~$T$~=~5.5 models are too close together to
distinguish between the two models.  Unfortunately, extended [O~III]
emission has not been detected in many of the galaxies, so our
analysis of the hybrid photoionization/TML models based on the
[O~III]/H$\alpha$ ratio is limited.  The extraplanar [O~III]/H$\alpha$
profiles in the five galaxies in which [O~III] has been detected fall
into two categories: three objects show an increasing
[O~III]/H$\alpha$ line ratio with increasing $\vert$z$\vert$, while
the [O~III]/H$\alpha$ ratio drops with height in the other two
galaxies.  The galaxies in the first category tend to be better fit by
a TML model with an intermediate mixing temperature (log~$T$ = 5.3),
while those in the second category are better fit by a model with
lower mixing temperatures (log~$T$ = 5.0).

For five of the galaxies in our sample, we find that the influence of
the TML regions on the emission-line spectra increases with height,
such that the contribution to the observed line ratios ({\em not} to
the line flux) increase roughly from $\sim$ 30 to $\sim$ 75\%.  These
results are consistent with models in which turbulent mixing occurs at
higher elevations, where superbubbles break out of the thin disk
layer, or at the locations where cooling halo gas is mixing with the
ejected gas.  There are exceptions to this general rule, however.  In
UGC~2092 and NGC~4013, the TML model contributes to the observed line
ratios a near constant amount (70\% and 40\%, respectively), while in
NGC~4217, the importance of the TML model appears to decrease with
height (from 40\% to 20\%).  Finally, one of the galaxies in our
sample (NGC~4302) demonstrates only little deviations from the
photoionization model, and these deviations are not well explained by
any of the TML models.

\subsubsection{\em Shocks}

Shull and McKee (1979) model interstellar radiating shocks such as
those due to supernovae events.  They include the ionizing effect of
the UV precursor in the preshock gas in an effort to make their models
self-consistent.  In their study, Shull and McKee discuss ten models.
In seven of the models they choose as their standard preshock
conditions a hydrogen particle density of 10 cm$^{-3}$, a magnetic
field B$_{0\perp}$ = 1$\mu$G perpendicular to the flow, and cosmic
metal abundances.  They vary the shock velocity from 40 to 130 km
s$^{-1}$.  In the other three, they fix the shock velocity at 100 km
s$^{-1}$ and vary, in turn, the density, magnetic field, and
metal abundances.

In the middle column of Figure~6, the predicted line ratios from the
shock models of Shull \& McKee are compared with the predictions from
the photoionization model (Sokolowski 1994). The solar abundance shock
models are represented by the open triangles (the solid triangle
represents Shull and McKee's shock model with depleted abundances),
while the photoionization models are represented by points joined by a
thin solid line. The hybrid photoionization/shock model that best fits
the data is shown as the dark solid line.  The models of Shull \&
McKee with shock velocities near 100 km s$^{-1}$ do well in predicting
the high [O~III]/H$\alpha$ ratios detected in most eDIG and the
relatively small [O~I]/H$\alpha$ ratios.  The biggest problem arises
when trying to model the galaxies which show a decreasing
[O~III]/H$\alpha$ line ratio with increasing $\vert$z$\vert$ (recall
that two out of five galaxies with detected extraplanar [O~III] show
this behavior).  In these galaxies, the [O~III]/H$\alpha$ ratio drops
below that which is predicted by the pure photoionization model.
Since all of the shock models predict enhanced [O~III]/H$\alpha$ line
ratios, they are unable to explain the negative [O~III]/H$\alpha$
vertical gradients.

In galaxies without extraplanar [O~III] data, we have to rely heavily
on the [S~II]/H$\alpha$ {\em vs.}  [N~II]/H$\alpha$ plot to determine
the best fitting models.  The line ratios for NGC~3628 and NGC~4013
can hardly be explained by a combination of shocks and photoionization
(shocks can contribute at most $\sim$ 10\% to the observed line
ratios).  In the other galaxies, the best-fitting model appears to be
the depleted abundances model with a shock velocity of 100 km
s$^{-1}$.  In general, the influence of the shocks on the observed
line ratios appears to increase with increasing heights (from $\sim$
10 -- 20\% near the disk of the galaxies up to $\sim$ 40 -- 60\% at
higher $\vert$z$\vert$).  In at least three of the galaxies, the
hybrid photoionization/shock model seems to imply that the ionization
parameter decreases from a value of log~$U$ = --3 near the plane of
the galaxy down to log~$U$ = --4 at heights $\vert$z$\vert$ $\ga$
1~kpc.

\subsubsection{\em Shock + Precursor}

In a pair of papers, Dopita and Sutherland (1995, 1996) present
a grid of models of low-density,  high-velocity photoionizing radiative
shocks.  Dopita and Sutherland model photons which propagate both
upstream through the preshock gas as well as downstream into the
recombination region of the shock.  Their grid spans a broad range in
shock velocity and magnetic parameter, $B/n_e^{1/2}$. They assume
solar abundances and a low-density radiative steady flow shock.  The
importance of the magnetic parameter comes into play because they
assume that the magnetic field is frozen into the flow.  Dopita and
Sutherland present models both with and without precursor regions.

In these models, an increase in the shock velocity causes an
increase in the ionization parameter in the pre-shock gas.  Since the
postshock plasma is the source of ionizing photons in radiative
shocks, the faster the shock the larger the flux of ionizing photons,
and hence the higher the ionizing parameter.  Dopita and Sutherland
also found that an increase in the magnetic parameter increases the
ionization parameter in the photoionization-recombination zone of the
shock, since the electron density is inversely proportional to the
magnetic parameter, and the ionization parameter is inversely
proportional to the electron density.

The predictions from Dopita and Sutherland's shock models are shown in
the right column of Figure~6. The grid spans a range in shock velocity
from 100 to 500 km s$^{-1}$ (the 500 km s$^{-1}$ line is
represented by a dashed line) and a range in the magnetic parameter
B/$n_e^{1/2}$ from 0 to 4 $\mu$G cm$^{3/2}$ (the 4 $\mu$G cm$^{3/2}$
line is represented by a dotted line).  Due to the large parameter
space represented by the Dopita and Sutherland shock models, it is
difficult to say with much certainty the exact parameters that are
needed to reproduce the observed line ratios.  In general, it appears
that models combining photoionization and shocks with no precursor do
better in reproducing the line ratios than those using the shocks +
precursor models. The measured ratios also seem to indicate an
increase in the shock velocity with increasing height.

\section{Summary}

Deep long-slit spectra reaching flux levels of a few 10$^{-18}$ erg
s$^{-1}$ cm$^{-2}$ arcsec$^{-2}$ were obtained of nine nearby, edge-on
spiral galaxies.  H$\alpha$, [N~II] $\lambda$6583, and [S~II]
$\lambda\lambda$6716, 6731 are detected out to a few kpc in all of
these galaxies.  Several other fainter diagnostic lines are also
detected over a slightly smaller scale.  The relative strengths,
centroids, and widths of the various emission lines provide
constraints on the electron density, temperature, reddening,
kinematics, and possible source(s) of ionization of the eDIG.  The
main conclusions from the analysis are the followings:

\begin{itemize}

\item[$\bullet$] Seven of the nine galaxies in the sample show a
general increase in the [N~II]/H$\alpha$ and [S~II]/H$\alpha$ line
ratios with increasing height. The extraplanar [N~II]/H$\alpha$ line
ratios reach values in excess of 1.5 in some galaxies.  Extraplanar
[O~III] line emission has been reliably detected in five galaxies, and
three of those show an increasing [O~III]/H$\alpha$ line ratio with
height.  Extraplanar [O~I] line emission has been detected in three of
the galaxies, and in each case, the [O~I]/H$\alpha$ line ratio
increases with increasing $\vert$z$\vert$.  These trends in the line
ratios are similar to those observed in other galaxies including our own.

\item[$\bullet$] He~I $\lambda$5876 is detected in the midplane of
four of the galaxies, with values ranging from 0.018 to 0.052.  This
suggests an effective stellar temperature range of 36,500 -- 38,500~K
and an upper limit to the stellar mass function of 42 -- 54
M$_{\odot}$.  Extraplanar He~I emission is detected in only one
galaxy, NGC~2820. The He~I/H$\alpha$ and [N~II]/H$\alpha$ line ratios
in the eDIG of this galaxy are consistent with O-star photoionization,
although the other ratios in this galaxy suggest the presence of
an additional source of ionization.

\item[$\bullet$] For all but one galaxy in the sample (NGC~4302),
photoionization by massive OB stars does not appear to be sufficient
to explain the line ratios in the eDIG. At least one other source of
ionization appears to be needed.  Hybrid models that combine
photoionization by OB stars and photoionization by turbulent mixing
layers (TMLs) or shocks provide the best fit to the data. In contrast
to the pure photoionization models, these hybrid models are able to
reproduce the run in the [O~III]/H$\alpha$ line ratios observed in
many of the galaxies.  Overall, the hybrid photoionization/TML models
do a better job of explaining the observed line ratios than the
photoionization/shock models, although in many cases the
photoionization/shock models do almost or just as well.  The
contribution of the turbulent mixing layers (or shocks) to the
observed line ratios appears to increase with increasing height. These
results are consistent with models in which turbulent mixing and
shocks occur at higher elevations, where superbubbles break out of the
thin disk layer, or at the locations where cooling halo gas is mixing
with the ejected gas.

\item[$\bullet$] Three galaxies in the sample appear to show
significant vertical velocity gradients. In NGC~3628, the presence of
a prominent dust lane is affecting the velocity measurements and
producing large variations in excess of 100 km s$^{-1}$ within a range
in height of only 0.5 kpc. Monotonic velocity gradients of order 50 --
70 km s$^{-1}$ kpc$^{-1}$ appear to exist in NGC~2820 and NGC~4013 (a
gradient may also be present in NGC~4302, but the uncertainties are
large). These gradients can be explained if the rotational velocities
of the eDIG are lower in the eDIG than in the disk, as predicted by
the galactic fountain model. The upper limits on the velocity
gradients in the other galaxies ($\sim$ 30 -- 50 km s$^{-1}$ depending
on the galaxy) are not inconsistent with this model.

\end{itemize}

\acknowledgments

This work has benefitted from several discussions with
J. Bland-Hawthorn.  The authors thank the referee, Dr. Ren\'e
Walterbos, for several suggestions which significanlty improved this
paper.  SV is indebted to the California Institute of Technology and
the Observatories of the Carnegie Institution of Washington for their
hospitality, and is grateful for partial support of this research by a
Cottrell Scholarship awarded by the Research Corporation, NASA/LTSA
grant NAG 56547, and NSF/CAREER grant AST-9874973.  STM was also
supported in part by NSF/CAREER grant AST-9874973. This work has made
use of NASA's Astrophysics Data System Abstract Service and the
NASA/IPAC Extragalactic Database (NED), which is operated by the Jet
Propulsion Laboratory, California Institute of Technology, under
contract with the National Aeoronautics and Space Administration.

\vskip 1.0in
\centerline{Appendix: Notes on Individual Objects}
\vskip 0.2in

\vskip 0.2in
\centerline{UGC~2092} 
\vskip 0.2in

This is a highly inclined ($i$ = 86$^\circ$; Guthrie 1992) Scd galaxy
that lies at a distance of 72~Mpc (PBS). PBS detected a bubble-like
structure approximately 20$\arcsec$ NE in the plane of the galaxy.  We
positioned the slit so that it passes directly through this structure
(Fig. 1).  UGC~2092 is the most distant galaxy in the sample.  In
order to detect even the strongest emission lines, the data were
spatially binned by a factor of two, yielding a spatial scale of
$\sim$ 1.1~kpc per spatial element.  The TML models have difficulties
explaining the observed line ratios.  Higher mixing gas temperatures
seem to be favored, but if this were the case, then the TML model
would have to account for almost all of the observed emission on the
west side of the galaxy, while the east side would require an
ionization parameter smaller than --5.0.  The shock models do a better
job of explaining the observed line ratios.  A shock velocity of 100
km sec$^{-1}$ and depleted abundances appear to best model the shock
environment.  According to the [S~II]/H$\alpha$ vs. [N~II]/H$\alpha$
plot, shocks play a more important role on the east side of the
galaxy, contributing about 60\% to the observed line ratios, while on
the west side, only 15\% -- 45\% of the observed line ratios is
produced through shocks.  Dopita \& Sutherland's shock models also
suggest a stronger influence from shocks on the eastern side of the
galaxy.  The observed line ratios can best be modeled by emission
originating in the recombination region of the shock, with a magnetic
parameter of $\sim$ 0 $\mu$G cm$^{3/2}$.

\vskip 0.4in
\centerline{UGC~3326}
\vskip 0.2in

UGC~3326 is a Scd galaxy which lies at a distance of 48~Mpc (PBS) with
an inclination angle of 90$^\circ$ (Guthrie 1992).  PBS detected a
faint plume on the SW side of the galaxy.  The plume extends outward
at an angle of 52$^\circ$ away from the axis of the galaxy out to a
height $\sim$ 1.3 kpc.  At the end of the plume is a relatively bright
cloudlike structure.  The slit was centered on the galaxy and ran
directly through this plume (Fig. 1).  UGC~3326 also required 2x
spatial binning in order to detect even the strongest emission lines,
yielding a spatial scale of $\sim$ 700~pc per spatial element.  As in
the case of UGC~2092, only [N~II]/H$\alpha$ and [S~II]/H$\alpha$ are
available to estimate the influence of a secondary source of
ionization.  The observed line ratios near the disk of the galaxy (out
to $\sim$~700~pc) are consistent with the photoionization model and do
not require a secondary ionization source.  At higher $\vert$z$\vert$
(on the northwest side of the galaxy), however, the [S~II]/H$\alpha$
line ratios are larger than what can solely be explained by the
photoionization model.  The TML model suggests that at higher
elevations, about 40 -- 80\% of the line ratios is produced by the
mixing layer gas, while the shock model suggests that 20 -- 40\% of
the line ratio arises from shocks (based on the 100 km sec$^{-1}$,
depleted abundances model).  Unfortunately, with only H$\alpha$,
[N~II] and [S~II] lines observed in this galaxy, we cannot further
constrain the data or verify which model best explains the observed
eDIG emission.

\vskip 0.4in
\centerline{UGC~4278}
\vskip 0.2in

R96 imaged UGC~4278, an SB(s)d galaxy with an inclination of
90$^\circ$ (Tully 1988), and detected prominent plumes on both sides
of the nucleus of the galaxy, out to a height of $\sim$ 1.4~kpc.  R96
suggests that this may trace outflow from a nuclear starburst.  Our
slit was positioned so it passes through the nucleus of the galaxy,
and therefore through the emission-line regions detected by R96
(Fig. 1).  UGC~4278 lies at a distance of 10.6~Mpc (Tully 1988),
yielding a spatial scale of $\sim$ 160~pc pixel$^{-1}$.  The
[O~III]/H$\alpha$ data in UGC~4278 tend to complicate the
identification of the secondary source of ionization.  While the data
for this line ratio, as well as [S~II]/H$\alpha$, suggest the need for
an additional source of ionization, they conflict in the determination
of exactly what that secondary source might be.  While the
[S~II]/H$\alpha$ line ratio shows a general increase with height, the
[O~III]/H$\alpha$ line ratio show a general decrease with height.
While the [S~II]/H$\alpha$ {\em vs.} [N~II]/H$\alpha$ data points may
be explained with a photoionization model with ionization parameters
from $\sim$ --2.0 to --3.0 mixed with shocked gas with a shock
velocity of 100 km sec$^{-1}$ in a depleted abundance region, neither
this model nor any other shock model can explain the decrease in the
[O~III]/H$\alpha$ line ratio with increasing height.  Similar problems
occur with the TML models. While the TML models with low mixing gas
temperatures (such as log~$T$~=~5.0) have negligible [O~III] emission,
and so could account for a decrease in this line relative to H$\alpha$
if TML becomes more important at large heights, the predicted
[S~II]/H$\alpha$ and [N~II]/H$\alpha$ line ratios for this model lie
close to the photoionization curve, and so limits the available
parameter space.  Looking at Figure 6, the observed [S~II]/H$\alpha$
and [N~II]/H$\alpha$ ratios are clearly larger than predicted by any
combinations of O-star photoionization and log~$T$ = 5.0 TML models.

The remarkably small [N~II]/H$\alpha$ line ratios in the disk of this
galaxy have been noted previously by Goad \& Roberts (1981).  In
their study of superthin galaxies, one of which was UGC~4278, they
derive a mean [N~II]/H$\alpha$ line ratio of 0.12 for this object,
consistent with our measurements.  They conclude that the small
[N~II]/H$\alpha$ ratios are a result of nitrogen deficiency in this
galaxy.  This suggests that star formation proceeds on a much longer
time scale than in most galaxies, since nitrogen is a secondary
product of nucleosynthesis. It also implies that large-scale shocks
are rare in UGC~4278.

\vskip 0.4in
\centerline{NGC~2820}
\vskip 0.2in

NGC~2820 lies at a distance of 20 Mpc, yielding a spatial scale of
$\sim$ 300~pc pixel$^{-1}$.  The position of the slit was selected
based on the narrowband images reported in Paper I. As discussed in \S
3.1.2, NGC~2820 is the only galaxy in our sample where extraplanar
He~I/H$\alpha$ emission has been detected out to $\vert$z$\vert$
$\sim$ 1~kpc.  It is also the only galaxy in which all four main line
ratios are detected out to large vertical heights, therefore providing
strong constraints on the nature of secondary ionization source.  The
main concern with NGC~2820 is that the majority of the
[O~III]/H$\alpha$ vs. [N~II]/H$\alpha$ data points lie below the
photoionization curve.  The TML models with low mixing gas
temperatures (log~$T$~=~5.0) also have low [O~III]/H$\alpha$ values.
Using this model in conjunction with the photoionization model, the
[O~III]/H$\alpha$ and [O~I]/H$\alpha$ data suggest that turbulent
mixing layers contribute roughly 35 -- 55\% to the observed line
ratios.  Unfortunately, the [S~II]/H$\alpha$ data do not support this
conclusion.  The [S~II]/H$\alpha$ data suggest a higher mixing gas
temperature (log~$T$~=~5.3), with the contribution to the line ratios
arising from this region apparently increasing with heights from 20 to
80\%.  The comparison with the photoionization/shock hybrid model
suggests that the data are best fit by the model with shock velocity
$v_s$ = 100 km sec$^{-1}$ and depleted abundance. With the exception
of those [O~III]/H$\alpha$ data points that fall below the
photoionization curve, the remaining data suggest that the ionization
parameter decreases from $\sim$ --3.0 to --4.0 while the contribution
of shocks to the line ratios rises from $\sim$ 10 to 40\%.  The Dopita
\& Sutherland models are perhaps more successful at explaining the
observed trends in the line ratios.  The [N~II]/H$\alpha$,
[S~II]/H$\alpha$, and [O~III]/H$\alpha$ line ratios lie in the region
of the (shock + precursor) model with a magnetic parameter of $\sim$ 0
$\mu$G cm$^{3/2}$.

\vskip 0.4in
\centerline{NGC~3628}
\vskip 0.2in

NGC~3628 is a starbursting Sb galaxy which is a member of the Leo
Triplet.  Located at a distance of 6.7~Mpc, it has an inclination
angle of 87$^\circ$ (Tully 1988).  Imaged by Fabbiano, Heckman, and \&
Keel (1990), they detected a large plume extending about 9~kpc from
the west side of the galaxy (see also Dahlem et al. 1996).  It is not
exactly perpendicular to the disk of the galaxy, but lies at a
position angle of 210$^\circ$ (while the disk of the galaxy lies at a
position angle of 104$^\circ$).  In order to cover as much of this
plume as possible, we positioned the slit to extend only to vertical
heights to the south side of the disk, following the extent of the
plume, rather than centered on the disk of the galaxy (see Fig. 1).
Fabbiano et al. also obtained spectroscopic information along the
minor axis of NGC~3628, though not through the detected plume.  They
detected a midplane [N~II]/H$\alpha$ ratio of 0.4 which increased to
$>$ 1.0 at higher $\vert$z$\vert$.  [S~II]/H$\alpha$ was also detected
and observed to increase with $\vert$z$\vert$, having a midplane value
of 0.3 and rising to $\sim$ 0.8 at large heights. Shock ionization was
proposed by Fabbiano et al. (1990) to explain these line ratios.
Figure 6 shows that pure shock or photoionization/shock hybrid models
are unable to reproduce the small [O~III]/H$\alpha$ line ratios in the
extraplanar plume; this is different from the conclusion of Fabbiano
et al (1990) who did not have constraints on the [O~III] emission. The
photoionization/TML hybrid models with a mixing gas temperature of
log~$T$~=~5.0 are more successful at explaining the observations.
Based on this model, the TML region would contribute roughly 50\% to
the observed line ratios, although the exact amount differs between
the [O~III]/$H\alpha$ and [S~II]/H$\alpha$ data points. The plume may
represent the interface where hot wind gas mixes with entrained disk
material (Dahlem et al. 1996).

\vskip 0.4in
\centerline{NGC~4013}
\vskip 0.2in

NGC~4013 is a Sb galaxy at a distance of 17~Mpc (260~pc pixel$^{-1}$;
Tully 1988) and with an inclination of 90$^\circ$ (Bottema 1995).  R96
detects extraplanar gas features extending up to $\vert$z$\vert$
$\sim$ 2.5~kpc on the northeast side, as well as a less prominent
line-emitting region on the southwest side.  Our recent imaging study
(Paper I) confirms the presence of eDIG in this object. We positioned
our slit to cover the northeast region.  The photoionization/TML
hybrid model with log~$T$~=~5.0 works well for NGC~4013.  While there
are only a few [O~III]/H$\alpha$ data points, they are consistent with
the [S~II]/$H\alpha$ data points, suggesting an ionization parameter
of $\sim$ --3.8 and implying that approximately 50\% of the line
ratios originate from the TML region.  The photoionization/shock
hybrid model is unable to reproduce the observed emission-line ratios.

\vskip 0.4in
\centerline{NGC~4217}
\vskip 0.2in

NGC~4217 is a Sb galaxy at a distance of 17~Mpc (260 pc pixel$^{-1}$;
Tully 1988) with an inclination of 86$^\circ$ (R96).  R96 detects two
bright plumes of extraplanar emission extending $\sim$ 2~kpc from
either side of the disk on the southwest side of NGC~4217.  Our slit
was positioned so that it passes through both plumes of extraplanar
gas.  [O~I] was detected out to $\sim$ 1.3~kpc on either side of
NGC~4217.  Based on the [O~I]/H$\alpha$ line ratios, in addition to
the observed [S~II]/H$\alpha$ line ratios, we find that the
photoionization/TML hybrid models predict that the ionization
parameter decreases from --4.0 to --5.0, while the influence of TMLs
on the observed line ratios remains relatively constant at $\sim$
40\%.  The best fit photoionization/shock model suggests an ionization
parameter which decreases from --3.5 to --4.8, and a shock model with
a velocity of 100 km sec$^{-1}$ and depleted abundances, which
provides only 10 to 20\% of the observed line ratios.  According to
the Dopita \& Sutherland shock models, the observed emission is
inconsistent with shock + precursor emission.  Although there is
significant scatter in the data, the observed line ratios appear to be
well described by emission originating from a region with a magnetic
parameter $<$ 1 $\mu$G cm$^{3/2}$ and a shock velocity which increases
with increasing height (up to almost 400 km sec$^{-1}$).

\vskip 0.4in
\centerline{NGC~4302}
\vskip 0.2in

NGC~4302 was imaged by both R96 and PBS.  It is an Sc galaxy at a
distance of 16.8~Mpc (250~pc pixel$^{-1}$; Tully 1988) with an
inclination of 90$^\circ$ (R96).  PBS detected an extremely faint
emission-line structure emerging 0.73 kpc from the plane close to the
nucleus of the galaxy.  R96 detects faint, diffuse extraplanar
emission at galactocentric radii $<$ 4~kpc up to $\vert$z$\vert$
$\sim$ 2~kpc.  Therefore, we chose to place our slit through the
center of the galaxy to incorporate both structures.  Unfortunately,
the uncertainties on many of the extraplanar line ratios are large and
prevent us from making strong statements on the source of ionization
in the eDIG of NGC~4302. All of the data points are consistent with
the photoionization model, within the large measurement errors.

NGC~4302 was also observed spectroscopically by Collins \& Rand
(2001).  The position of their slit is 20$\arcsec$ southeast of ours,
or given a distance of 16.8~Mpc, $\sim$ 1.6~kpc southeast.  They
detect the prominent [N~II] and [S~II] lines up to $z$ = 2~kpc on
either side of the galaxy, but poor weather conditions prohibited the
detection of [O~III].  Their [N~II]/H$\alpha$ line ratios increase
from 0.4 up to 1.4 at $\vert z\vert$ = 1.5~kpc, and their detected
[S~II]/[N~II] line ratios remain constant at $\sim$ 0.6.  These values
suggest a gas temperature ranging from 6600 to 10,000~K.  While their
line ratios differ from ours, we do see the same general increase in
the emission line ratios with $\vert$z$\vert$ .

\vskip 0.4in
\centerline{NGC~5777}
\vskip 0.2in

In NGC~5777, an Sb galaxy with an inclination of 83$^\circ$ (Guthrie
1992) located at a distance of 25~Mpc (380~pc pixel$^{-1}$; PBS), PBS
detected a prominent extraplanar plume extending 1.1 kpc directly
north-east from a bright H~II region located on the southeast side of
the galaxy.  We positioned our slit so that it was centered on the
galaxy disk and passed through this plume.  A photoionization/TML
hybrid model with a mixing gas temperature of log~$T$~=~5.3 best fits
the data.  This fit suggests that the ionization parameter decreases
from --4.0 to --5.0 while the contribution from the TML region to the
emission line ratios increases from $\sim$ 20 to 80\%.  The
photoionization/shock hybrid model is less successful at explaining
the data. The best fit is found with a model with a shock velocity of
100 km sec$^{-1}$ and depleted abundances, where the contribution from
the shock models to the emission line ratios increases from $\sim$ 15
to 40\% with increasing heights.  The Dopita \& Sutherland models also
have difficulties reproducing the data. The [O~III]/H$\alpha$ ratios
measured at high $\vert$z$\vert$ favor a shock + precursor model,
while the [S~II]/H$\alpha$ data suggest that the emission originates
from shocks without precursor. These ratios also suggest different
values for the magnetic parameter.

\clearpage

\figcaption{Position of the slit superposed on the
continuum-subtracted H$\alpha$ image of (a) the NE side of UGC~2092,
(b) the SW side of UGC~3326, (c) UGC~4278, (d) NGC~2820, (e) NGC~3628,
(f) NGC~4013, (g) NGC~4217, (h) NGC~4302 and (i) NGC~5777 (see Table~1
for references to the images).  The
position of the slit is chosen to pass through at least one known
region of eDIG. In most cases the slit is positioned perpendicular to
the plane of the galaxy. In the others, it is positioned to maximize
coverage of the eDIG. UGC~2092 is rotated clockwise by 32$^\circ$ such
that the disk of the galaxy runs vertically, while UGC~3326 was
rotated clockwise by 84$^\circ$.  The other figures have not been
rotated (North is at the top of the figure and east to the left).}

\figcaption{Emission line spectra as a function of position along the
slit.  For display purposes, the spectra are scaled individually (or
in sets of two or three) with the scaling factor in relation to the
spectrum containing the H$\alpha$ + [N~II] complex listed in the upper
right corner of each spectrum.  The expected redshifted positions of
the emission lines as determined from the systemic velocity of the
galaxies are indicated by the dashed lines.  Tickmarks on the
horizontal axis are separated by 25 \AA\ in the observer's restframe.
In the cases where no line is detected, a representative
spectrum from near the disk of the galaxy is presented.  The presence
of the Na ID $\lambda\lambda$5897.6, 5891.6 absorption lines near He~I
is apparent in some of the galaxies (e.g., NGC~3628, NGC~4217,
NGC~4302 and UGC~3326).}

\figcaption{Vertical profiles of [N~II] $\lambda$6583/H$\alpha$, [S~II]
$\lambda$6716/H$\alpha$, [S~II] $\lambda$6716/[N~II] $\lambda$6583,
[O~III] $\lambda$5007/H$\alpha$, [O~I] $\lambda$6300/H$\alpha$, He~I
$\lambda$5876/H$\alpha$, H$\alpha$/H$\beta$, [S~II]
$\lambda$6716/[S~II] $\lambda$6731, and [O~III] $\lambda$5007/[O~III]
$\lambda$4959. This last ratio is expected to be constant ($\sim$ 3)
within the uncertainties. The error bars represent one-$\sigma$
uncertainties.  Also shown in each panel are the spatial profiles of the
H$\alpha$ emission (solid line) and continuum emission (dashed
line), both of which have been arbitrarily scaled for display
purposes.  Only ratios involving detected lines have been
included in this figure.}

\figcaption{Temperature profile and ionization fraction in the eDIG.
In the bottom left panel is a plot of the temperature profile (in
units of 10$^4$ K) based on the [N~II] $\lambda$6583/H$\alpha$ line
ratio presented in the upper left panel [see eqn. (2)].  Using this
temperature profile, [S~II] $\lambda$6716/H$\alpha$, [S~II]
$\lambda$6716/[N~II] $\lambda$6583, [O~III] $\lambda$5007/H$\alpha$
and [O~I] $\lambda$6300/H$\alpha$ have been calculated and
superimposed on the observed line ratios [see \S 4.2 and eqns. (3)
through (6) for details].  The lines represent different ionization
fractions ranging from 12.5\% (first solid line) to 100\% (last dotted
line).  The top two panels to the right of the [N~II]/H$\alpha$ plot
represent the S$^+$/S ratio; the bottom, middle plot represents the
O$^{++}$/O ratio; and the bottom, right plot represents the H$^+$/H
ratio. The [N~II]/H$\alpha$ ratios cannot be used to derive
the temperature profile in objects with strong eDIG [O~III]/H$\alpha$
ratios ($\ga$ 0.25); four objects are therefore excluded from the
analysis (UGC~4278, NGC~2820, NGC~4302, and NGC~5777; see text).}

\figcaption{Velocity and line width profiles. In three of the panels, the
velocity centroids of the H$\alpha$, [N~II] $\lambda$6583 and [S~II]
$\lambda$6716 emission lines are plotted as a function of position
along the slit.  The velocities are plotted relative to the systemic
velocity in km s$^{-1}$.  The dot-dashed line represents the average
position of the line centroid. One-$\sigma$ error bars have been
computed based on the accuracy in detecting the true line centroid as
a function of signal-to-noise ratio.  The bottom, right figure
presents the measured H$\alpha$ line widths in km s$^{-1}$ as a
function of vertical height. These measurements have been corrected
for the width of the instrumental profile.}

\figcaption{(a) Comparisons of observed [S~II] $\lambda$6716/H$\alpha$,
[O~III] $\lambda$5007/H$\alpha$, [O~I] $\lambda$6300/H$\alpha$, and
[N~II] $\lambda$6583/H$\alpha$ line ratios with the predictions of
photoionization, turbulent mixing layer, and shock models.  The data
for UGC~2092 are the filled circles with error bars.  The series of
points joined by a thin solid line in the panels in the left and
middle columns represent the predictions from photoionization models
of a hardened spectrum and dust depleted abundances with ionization
parameters (log~$U$) of --3.0, --3.5, --4.0, --4.2, --4.5, and --5.0
(note that [N~II]/H$\alpha$ decreases with increasing log $U$;
Sokolowski 1994).  The left column overlays the predictions from the
turbulent mixing layer (TML) models of Slavin et al. (1993).  These
values are derived from models with a hot gas velocity of 25 km
s$^{-1}$, depleted abundances, and an intermediate temperature after
mixing of log $T$ = 5.0, log $T$ = 5.3, or log $T$ = 5.5 (represented
by open triangles with size increasing with temperature). The hybrid
photoionization/TML model which best fits the data is shown as a thick
solid line. This model incorporates TMLs with a mixing temperature of
log~$T$ = 5.5 (see text for more detail). The middle column overlays
shock model predictions from Shull \& McKee (1979).  The values
plotted are calculated from solar abundance models with shock
velocities of 90, 100, and 110 km s$^{-1}$ (represented by open
triangles of increasing size).  A model with a shock velocity of 100
km s$^{-1}$ and depleted abundances (filled triangle) is also shown
for comparison.  For this galaxy as well as the others, the best fit
to the photoionization/shock hybrid model is derived using the shock
model with depleted abundances. The panels in the right column
presents the grid of predictions from the shock and shock+precursor
solar abundance models of Dopita \& Sutherland (1995).  The shock
velocity ranges from 100 to 500 km s$^{-1}$ (the 500 km s$^{-1}$ line
is represented by a dashed line) and the magnetic parameter
B/$n_e^{1/2}$ ranges from 0 to 4 $\mu$G cm$^{3/2}$ (the 4 $\mu$G
cm$^{3/2}$ line is represented by a dotted line).  (b) Same as Figure
6$a$ but for UGC 3326.  The best fit for the photoionization/TML
hybrid model uses a mixing temperature of log $T$ = 5.3.  (c) Same as
Figure 6$a$ but for UGC 4278.  The best fit for the
photoionization/TML model uses a mixing temperature of log $T$ = 5.0.
(d) Same as Figure 6$a$ but for NGC 2820.  The best fit for the
photoionization/TML hybrid model uses a mixing temperature of log $T$
= 5.3.  (e) Same as Figure 6$a$ but for NGC 3628.  The best fit for
the photoionization/TML hybrid model uses a mixing temperature of log
$T$ = 5.0.  (f) Same as Figure 6$a$ but for NGC 4013.  The best fit
for the photoionization/TML hybrid model uses a mixing temperature of
log $T$ = 5.0.  (g) Same as Figure 6$a$ but for NGC 4217.  The best
fit for the photoionization/TML hybrid model uses a mixing temperature
of log $T$ = 5.0.  (h) Same as Figure 6$a$ but for NGC 4302.  These
data do not require a secondary source of ionization.  (i) Same as
Figure 6$a$ but for NGC 5777.  The best fit for the
photoionization/TML hybrid model uses a mixing temperature of log $T$
= 5.3}

\clearpage

\setcounter{figure}{0}

\epsscale{1.0}
\begin{figure}[htbp]
\figurenum{1}
%\plotone{f1a.eps}
\caption{See jpg image.}
\end{figure}
\epsscale{1.0}

\epsscale{1.0}
\begin{figure}[htbp]
\figurenum{1}
%\plotone{f1b.eps}
\caption{(Cont'd) See jpg image.}
\end{figure}
\epsscale{1.0}

\epsscale{1.0}
\begin{figure}[htbp]
\figurenum{1}
%\plotone{f1c.eps}
\caption{(Cont'd) See jpg image.}
\end{figure}
\epsscale{1.0}

\clearpage

%\epsscale{0.4}
\begin{figure}[htbp]
\figurenum{2}
\plottwo{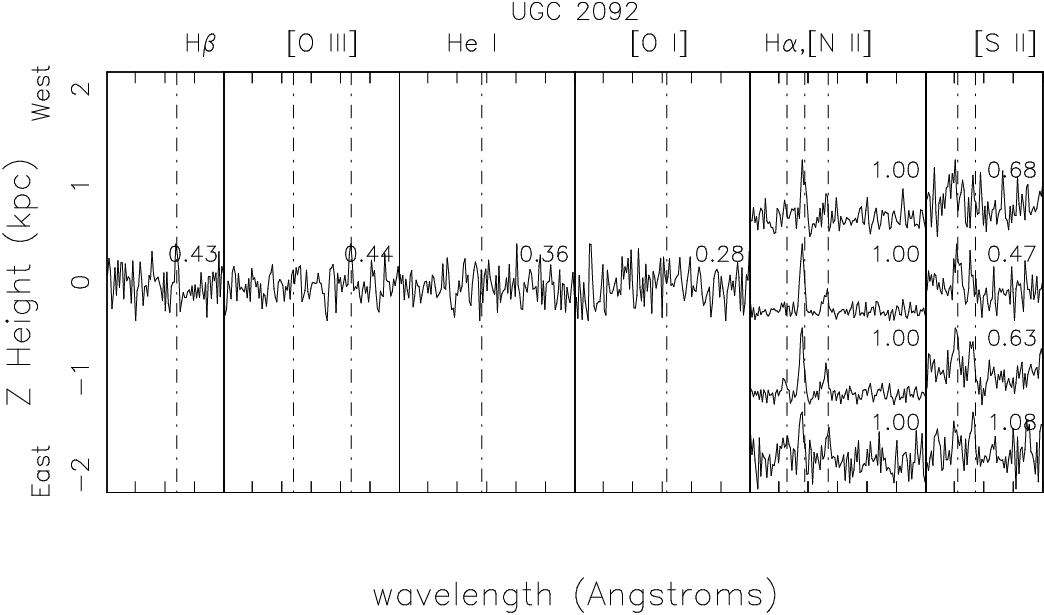}{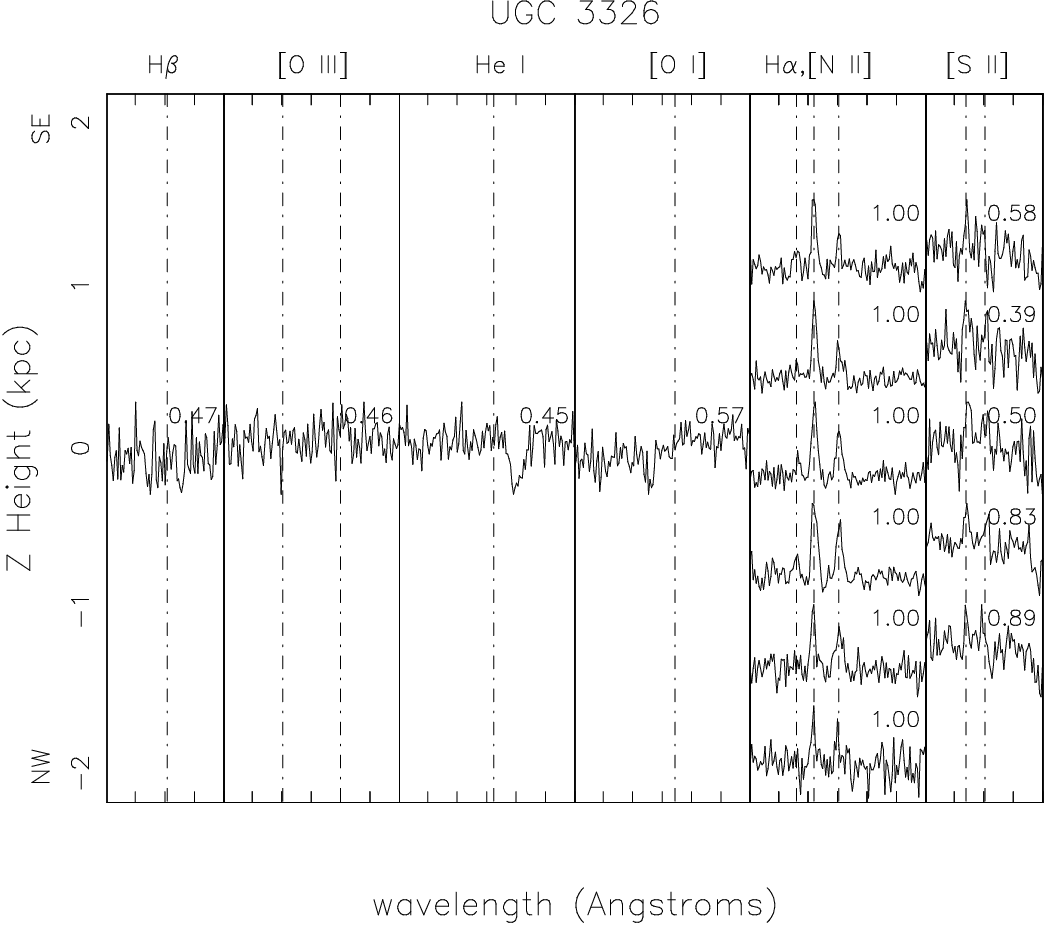}
\caption{}
\end{figure}
\epsscale{1.0}

%\epsscale{0.4}
\begin{figure}[htbp]
\figurenum{2}
\plottwo{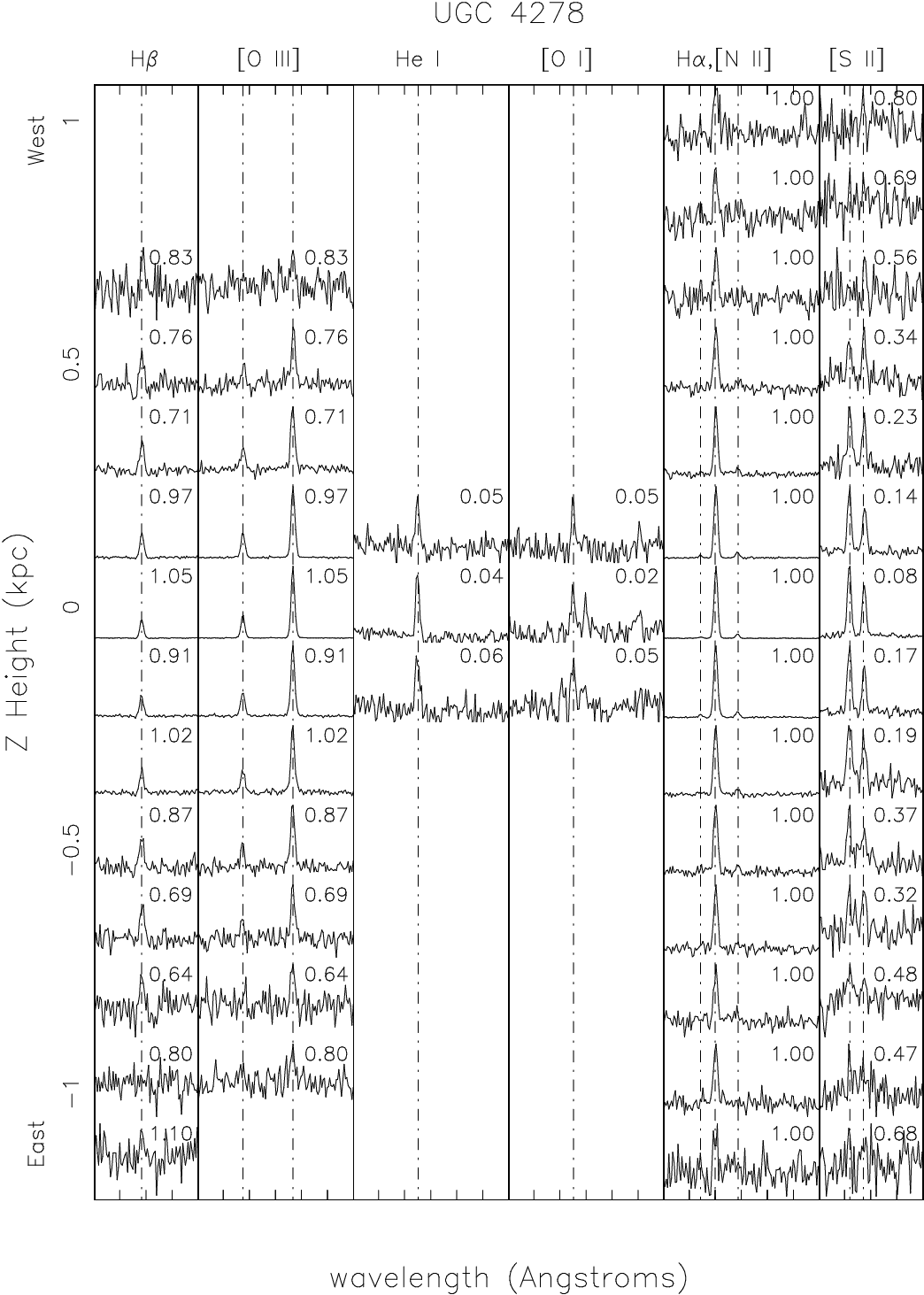}{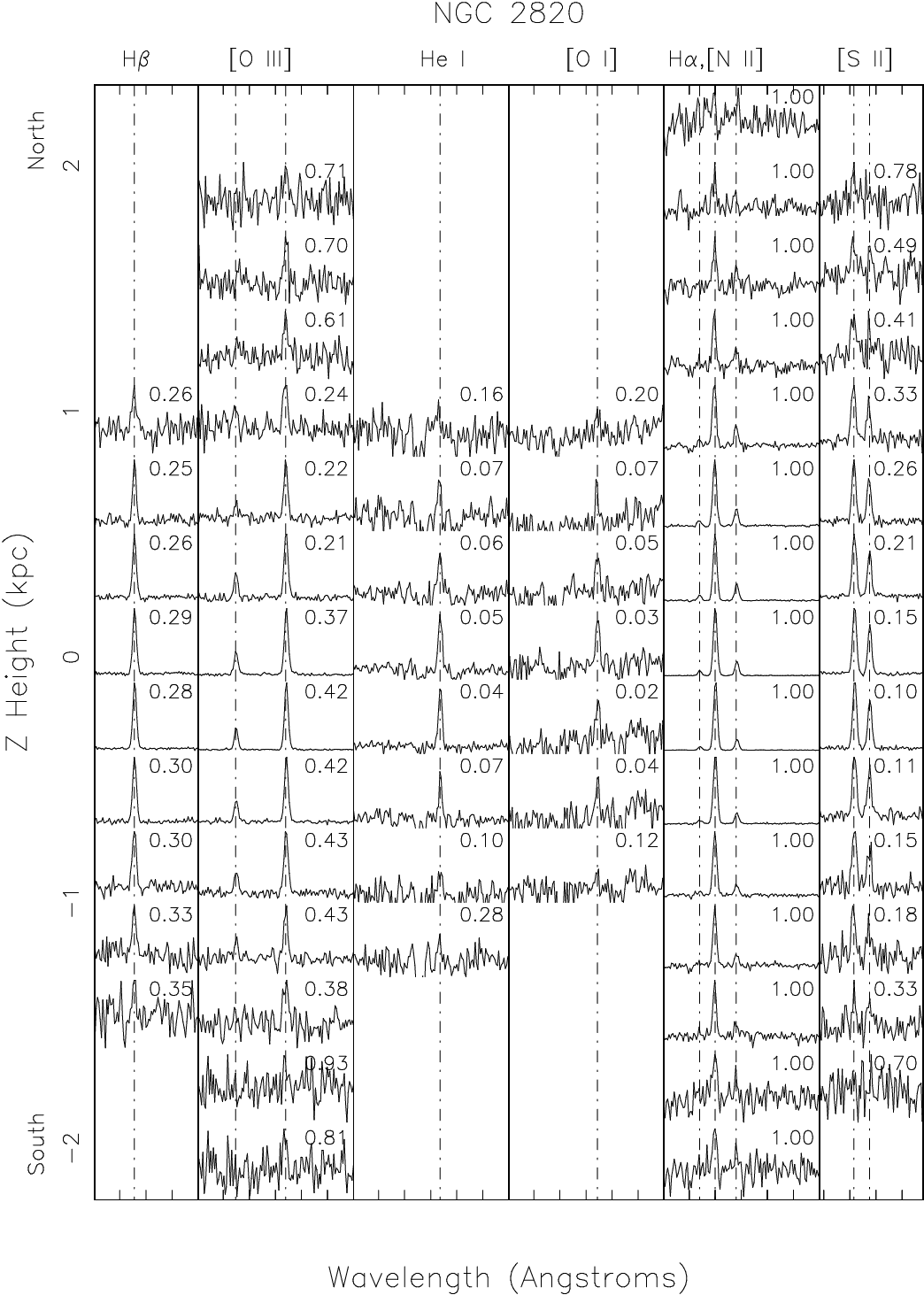}
\caption{(Cont'd.)}
\end{figure}
\epsscale{1.0}

%\epsscale{0.4}
\begin{figure}[htbp]
\figurenum{2}
\plottwo{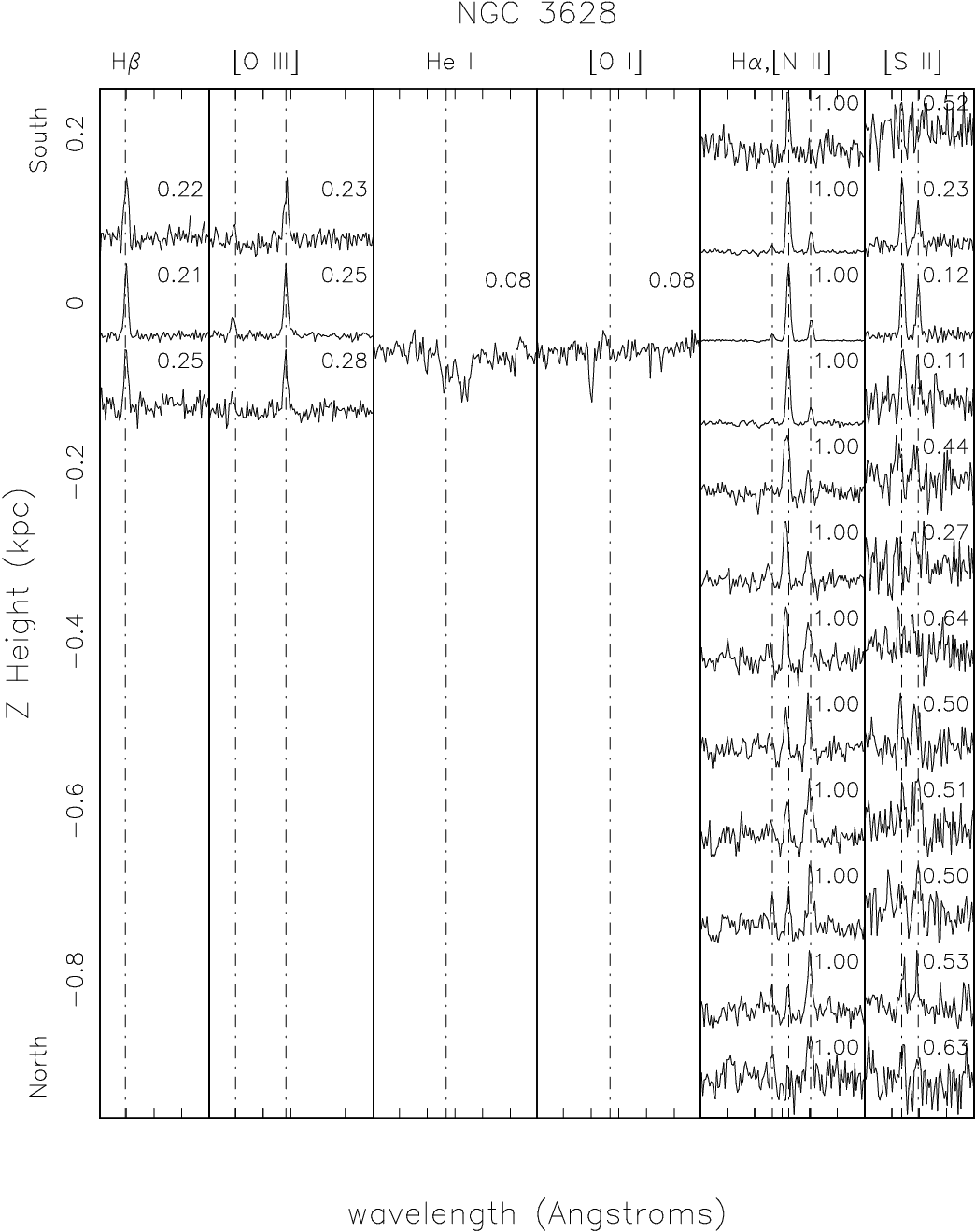}{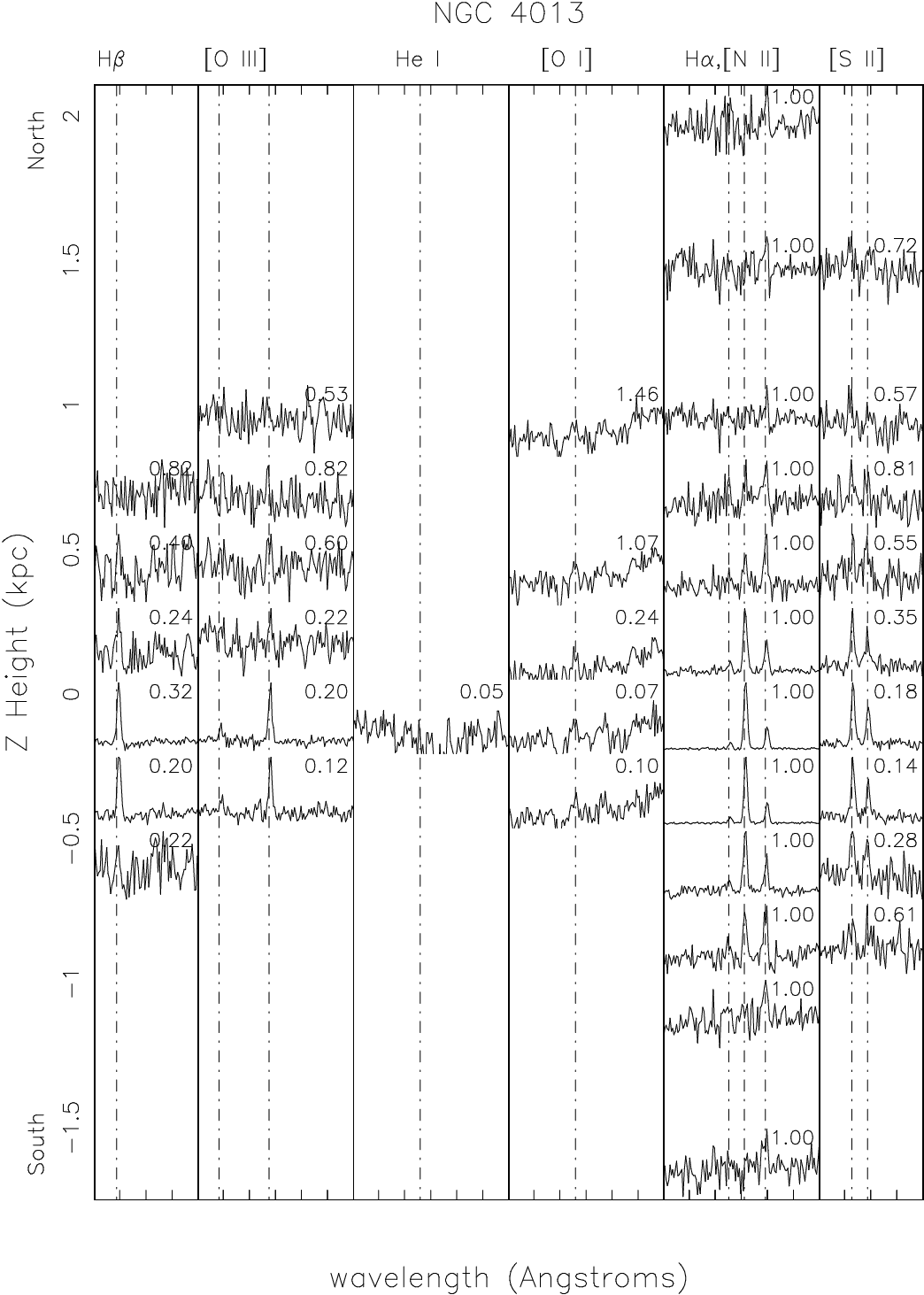}
\caption{(Cont'd.)}
\end{figure}
\epsscale{1.0}

%\epsscale{0.4}
\begin{figure}[htbp]
\figurenum{2}
\plottwo{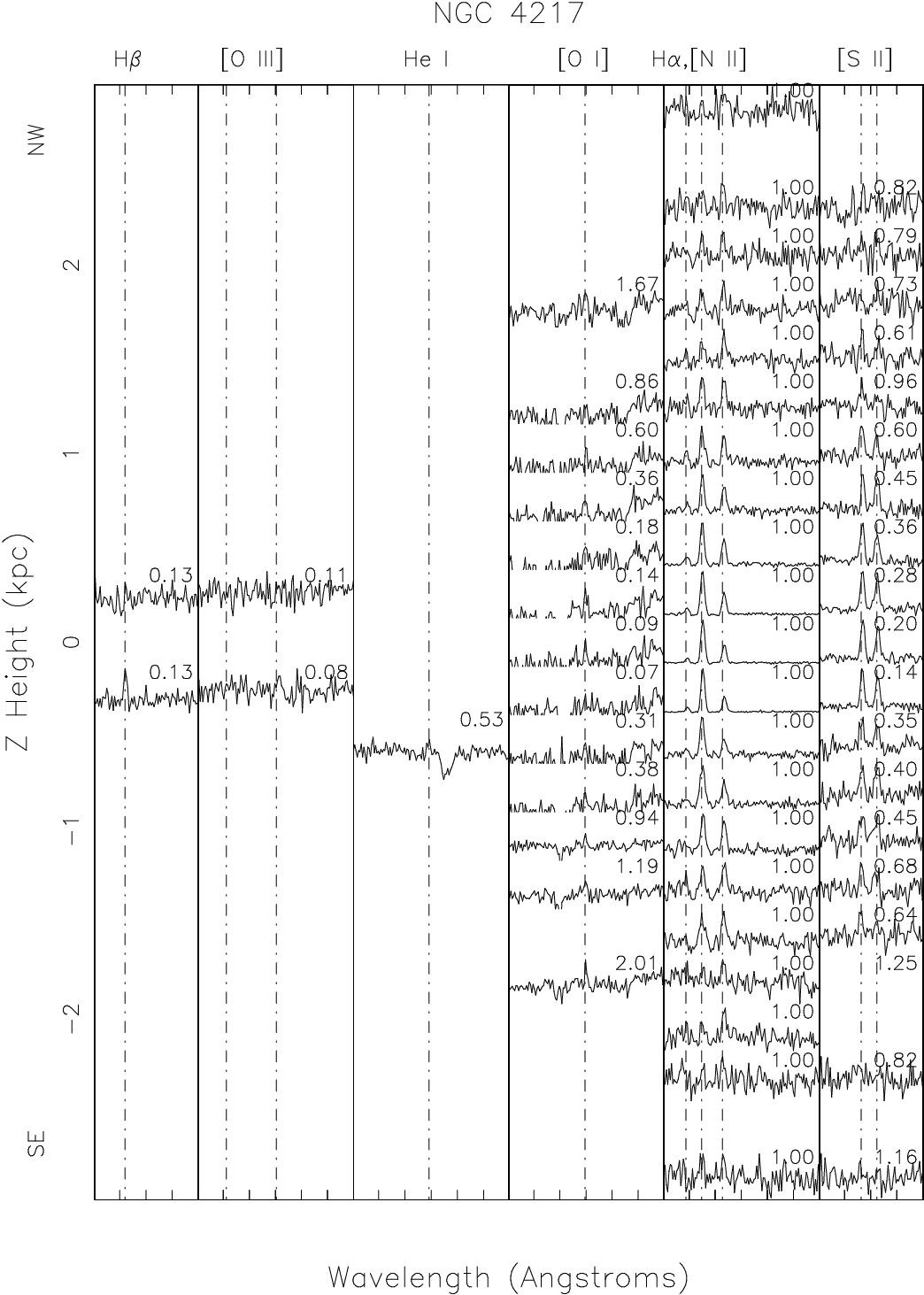}{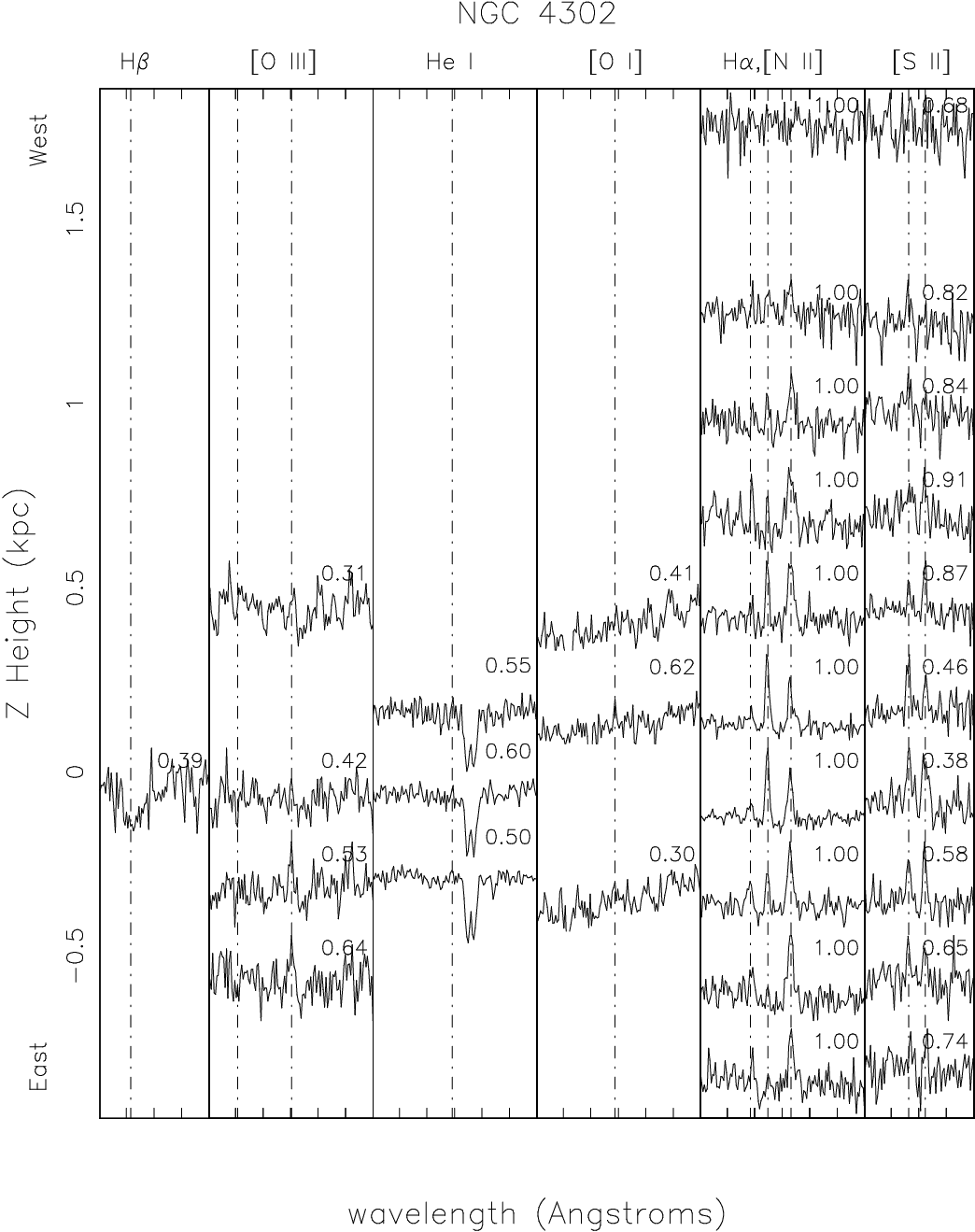}
\caption{(Cont'd.)}
\end{figure}
\epsscale{1.0}

%\epsscale{0.4}
\begin{figure}[htbp]
\figurenum{2}
\plottwo{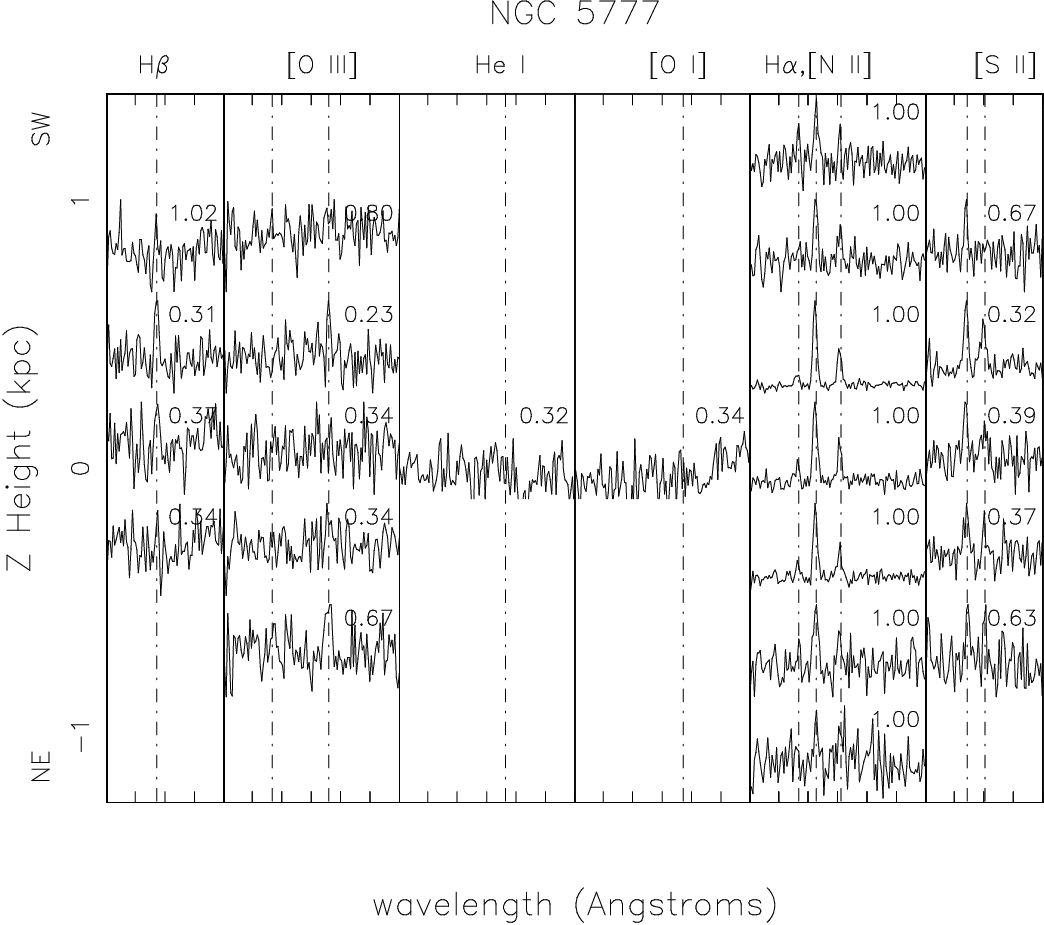}{f2i.eps}
\caption{(Cont'd.)}
\end{figure}
\epsscale{1.0}

\clearpage

\epsscale{0.65}
\begin{figure}[b]
\figurenum{3}
\plotone{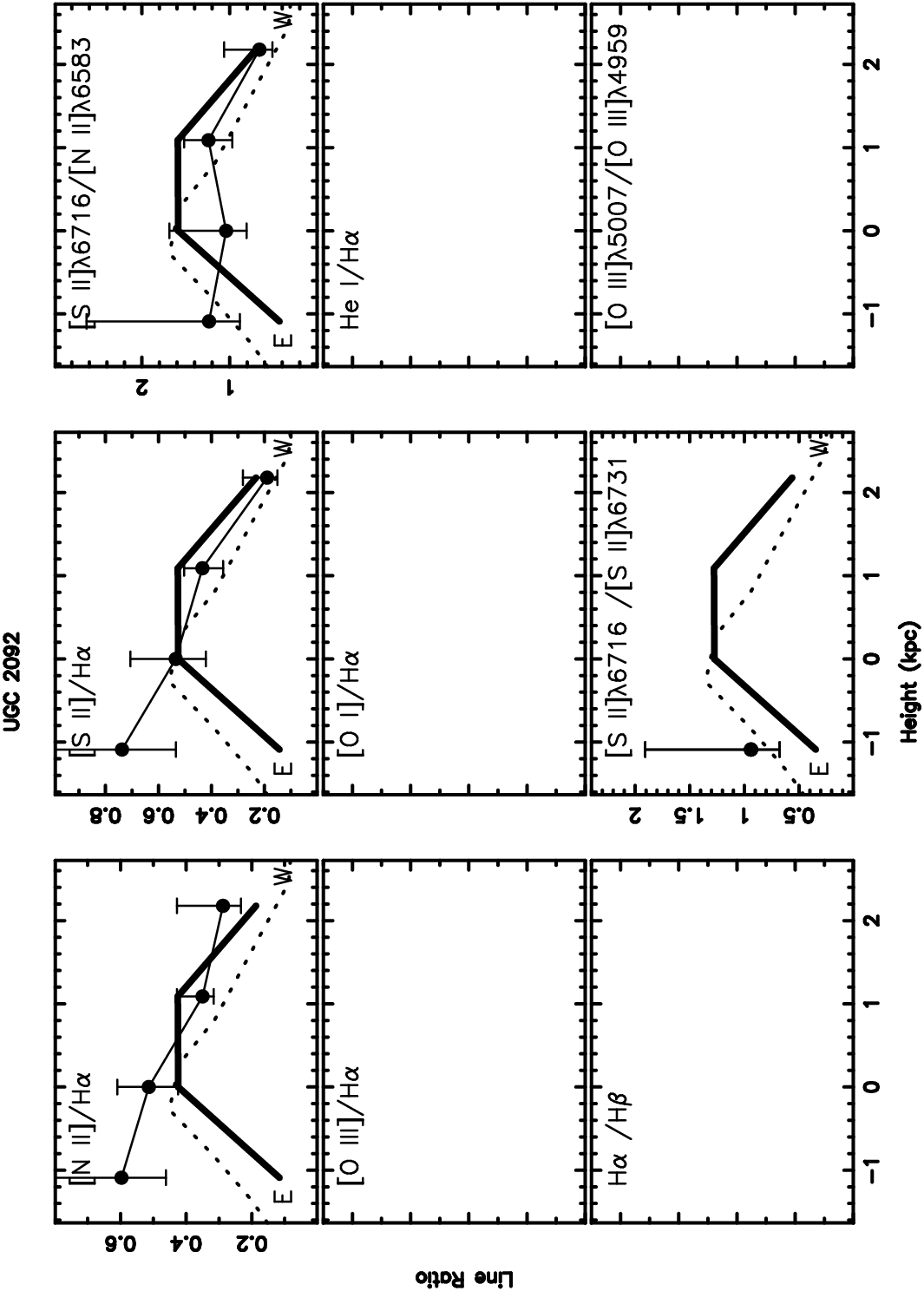}
\caption{}
\end{figure}
\epsscale{1.0}

\epsscale{0.65}
\begin{figure}[b]
\figurenum{3}
\plotone{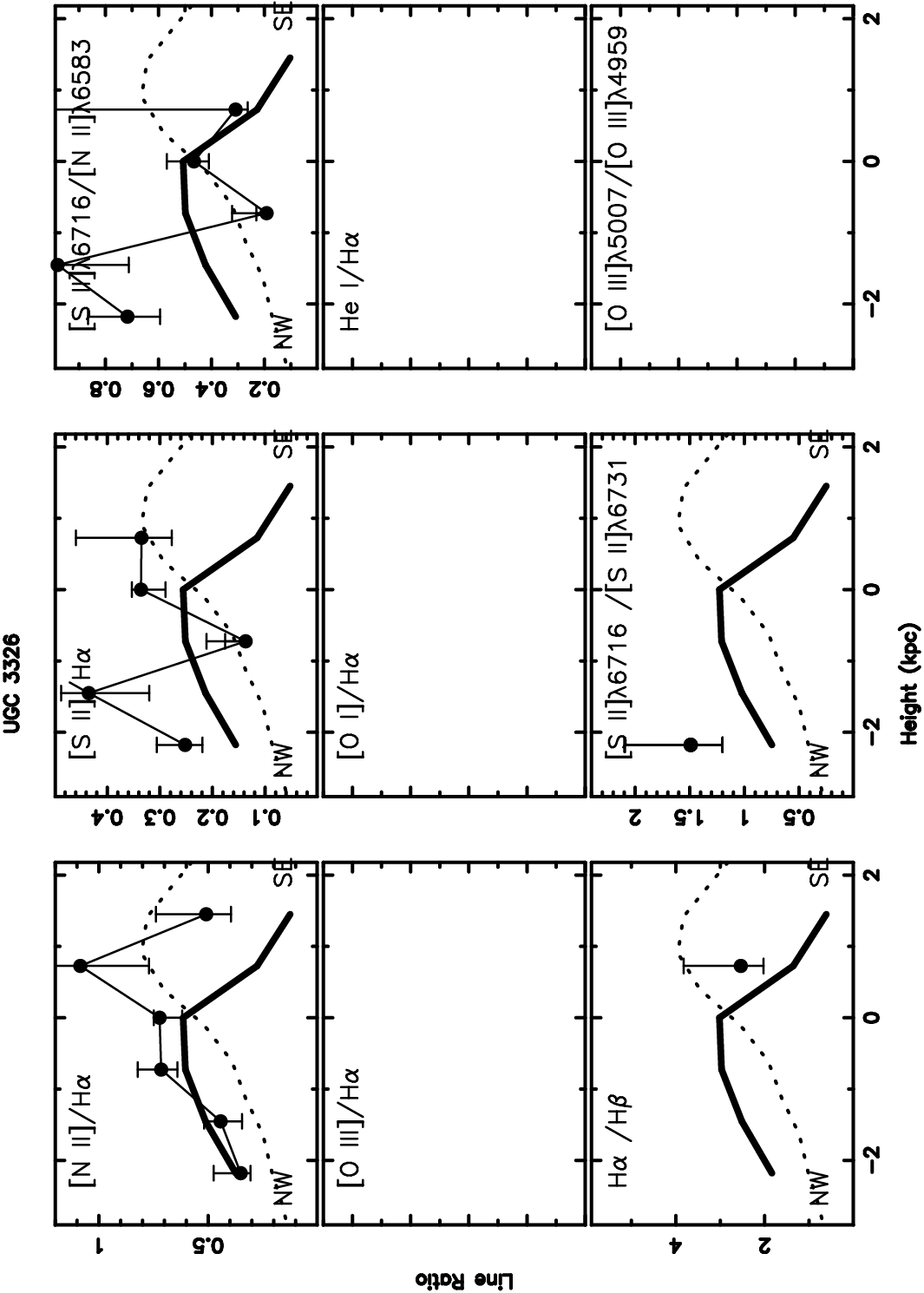}
\caption{(Cont'd.)}
\end{figure}
\epsscale{1.0}

\clearpage

\epsscale{0.65}
\begin{figure}[b]
\figurenum{3}
\plotone{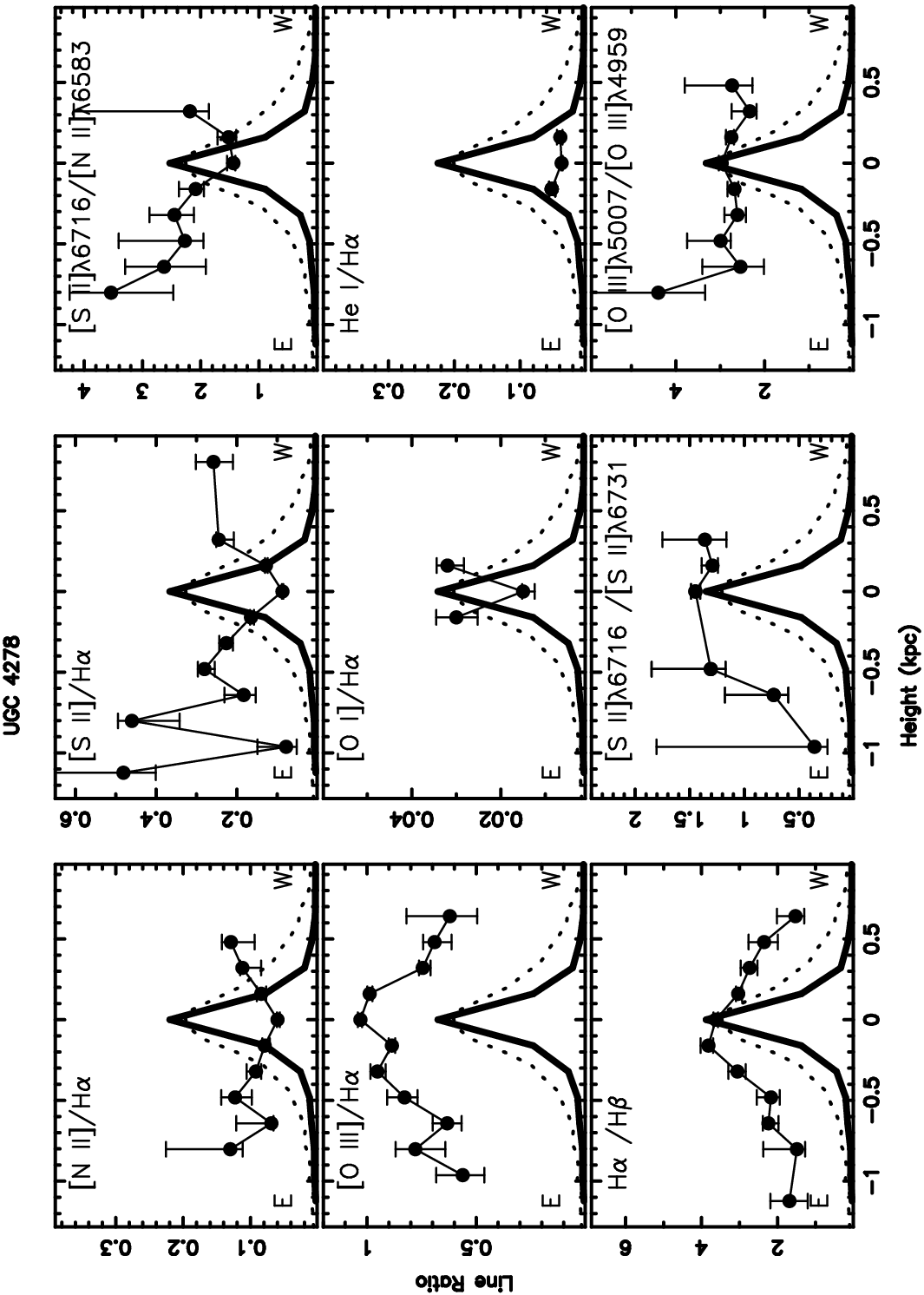}
\caption{(Cont'd.)}
\end{figure}
\epsscale{1.0}

\epsscale{0.65}
\begin{figure}[b]
\figurenum{3}
\plotone{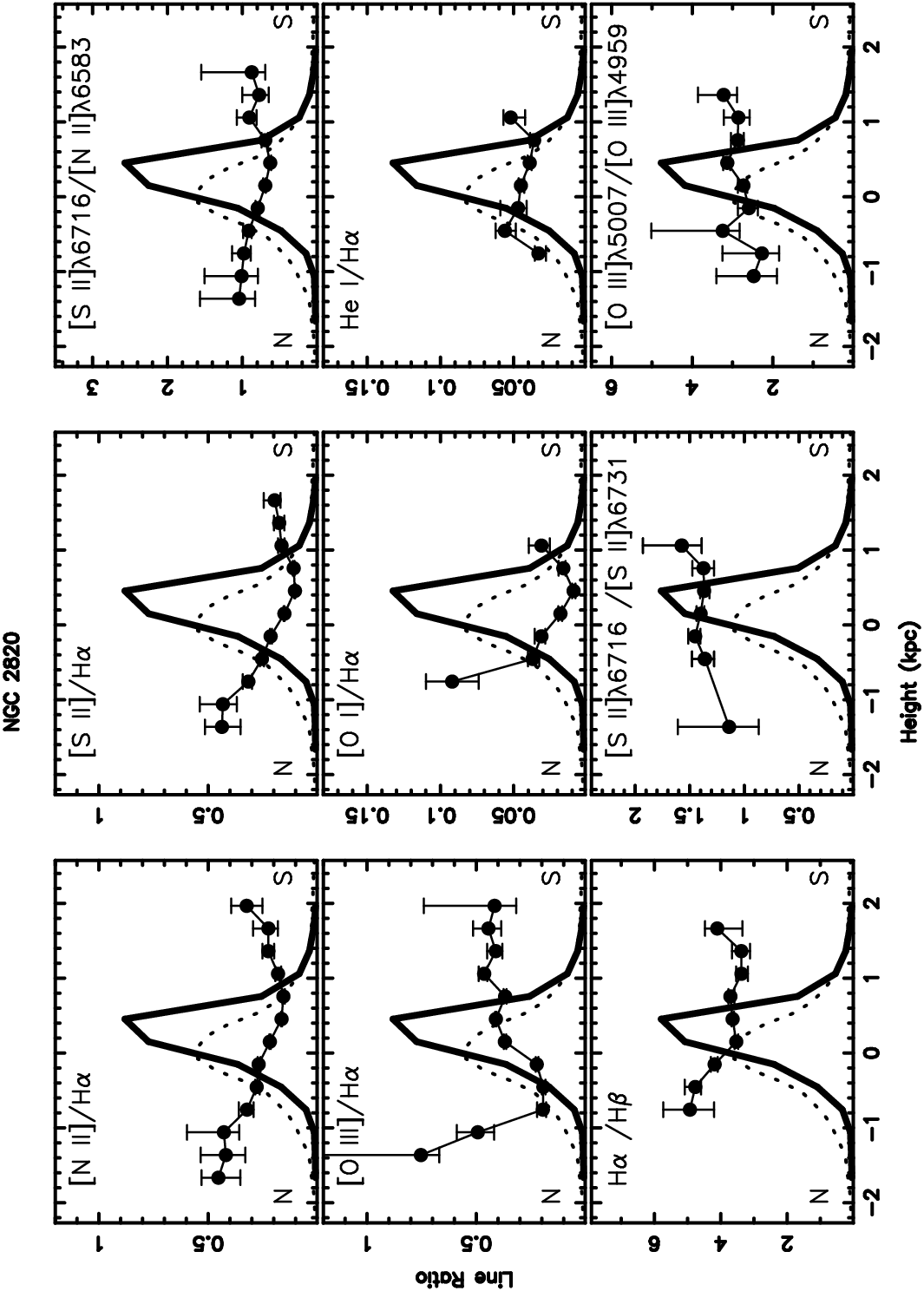}
\caption{(Cont'd.)}
\end{figure}
\epsscale{1.0}

\clearpage

\epsscale{0.65}
\figurenum{3}
\begin{figure}[b]
\plotone{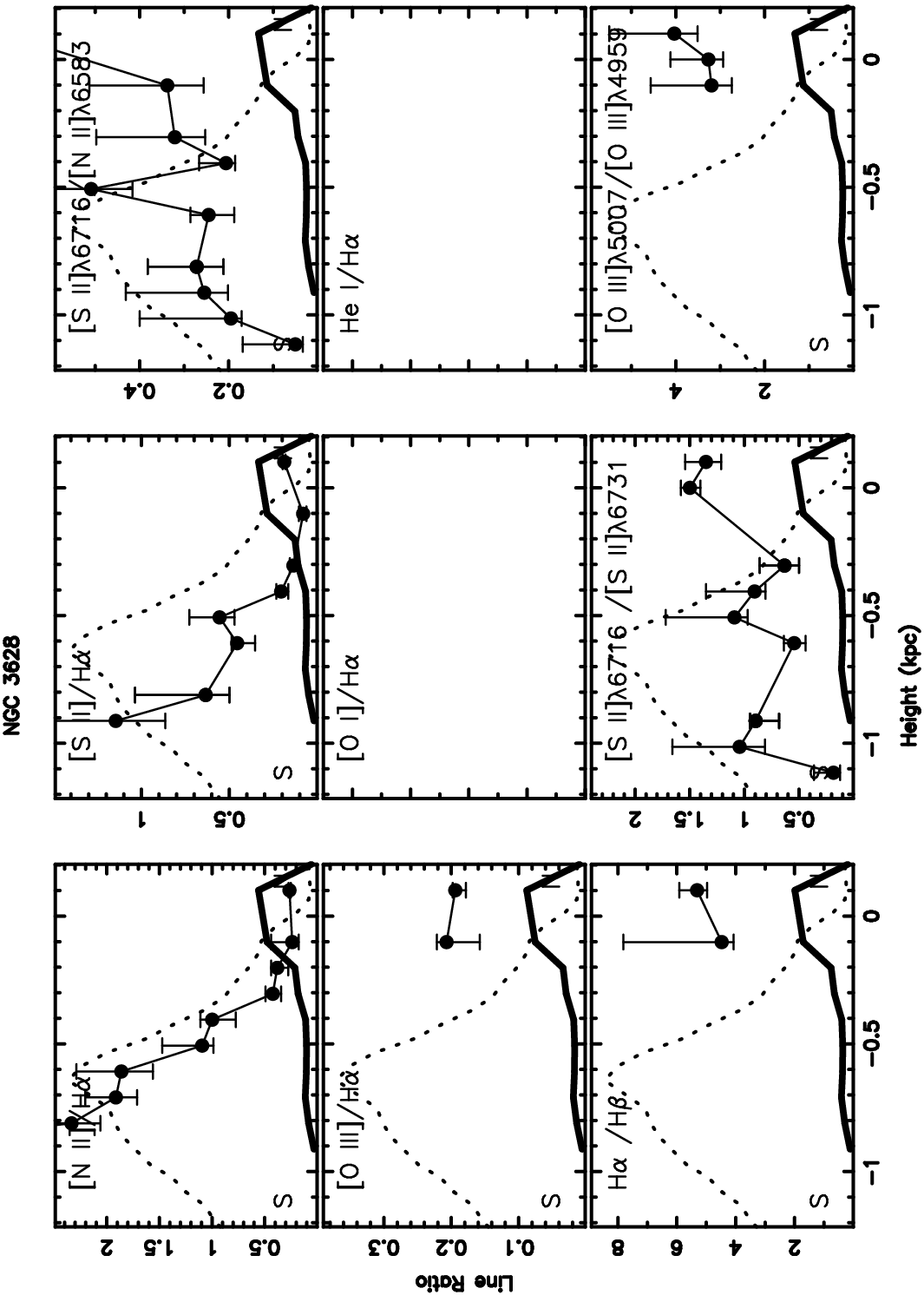}
\caption{(Cont'd.)}
\end{figure}
\epsscale{1.0}

\epsscale{0.65}
\begin{figure}[b]
\figurenum{3}
\plotone{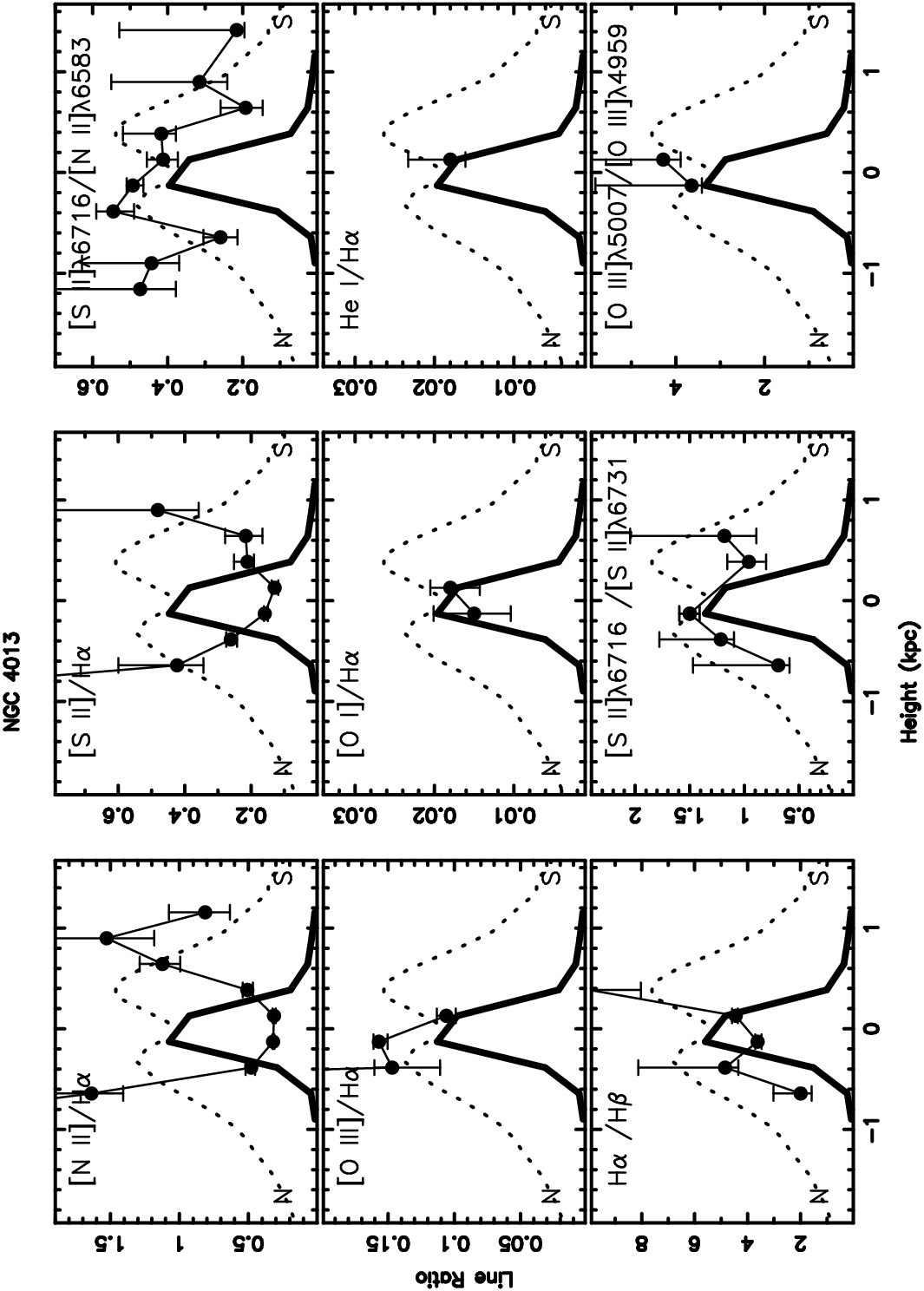}
\caption{(Cont'd.)}
\end{figure}
\epsscale{1.0}

\clearpage

\epsscale{0.65}
\begin{figure}[b]
\figurenum{3}
\plotone{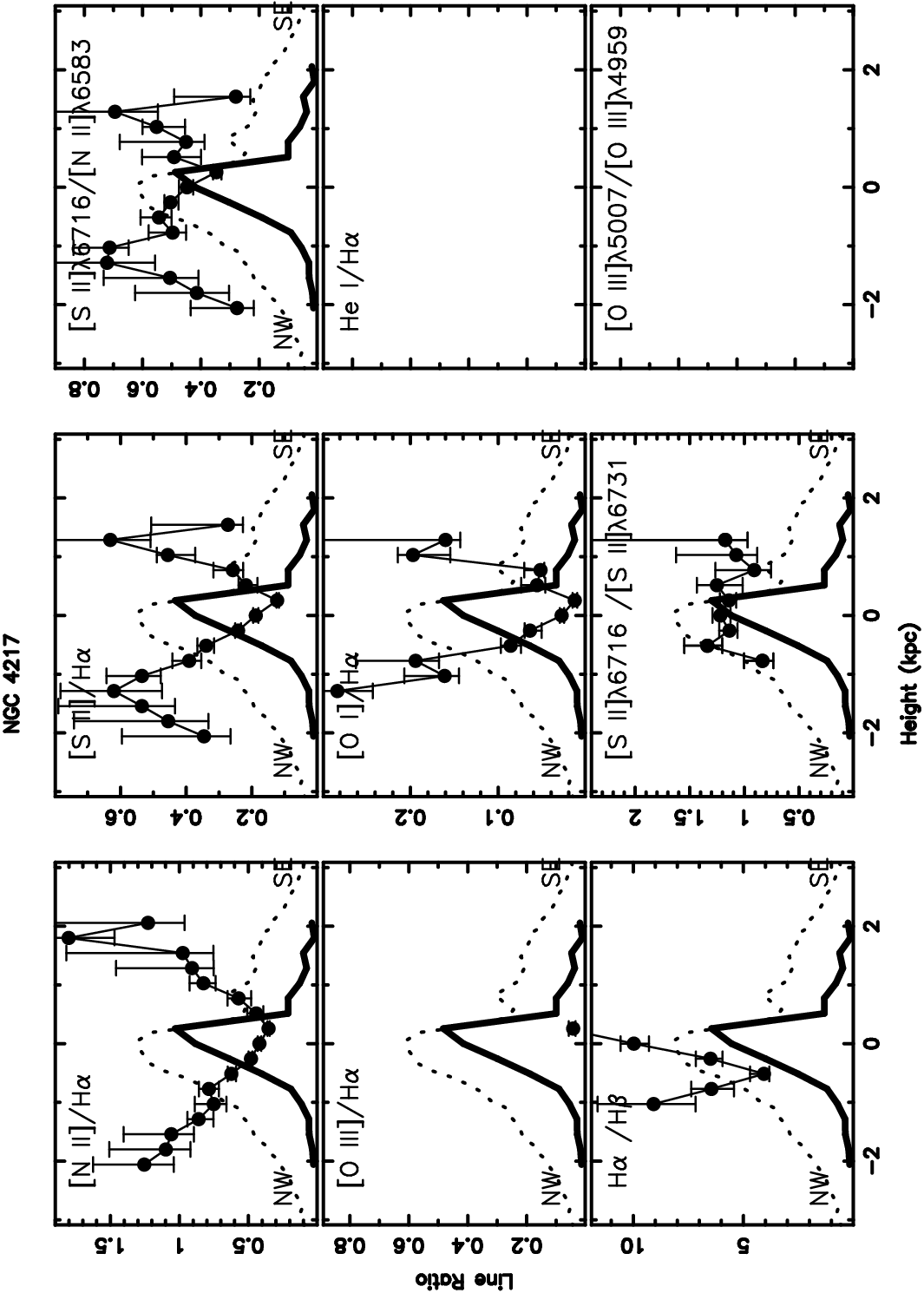}
\caption{(Cont'd.)}
\end{figure}
\epsscale{1.0}

\epsscale{0.65}
\begin{figure}[b]
\figurenum{3}
\plotone{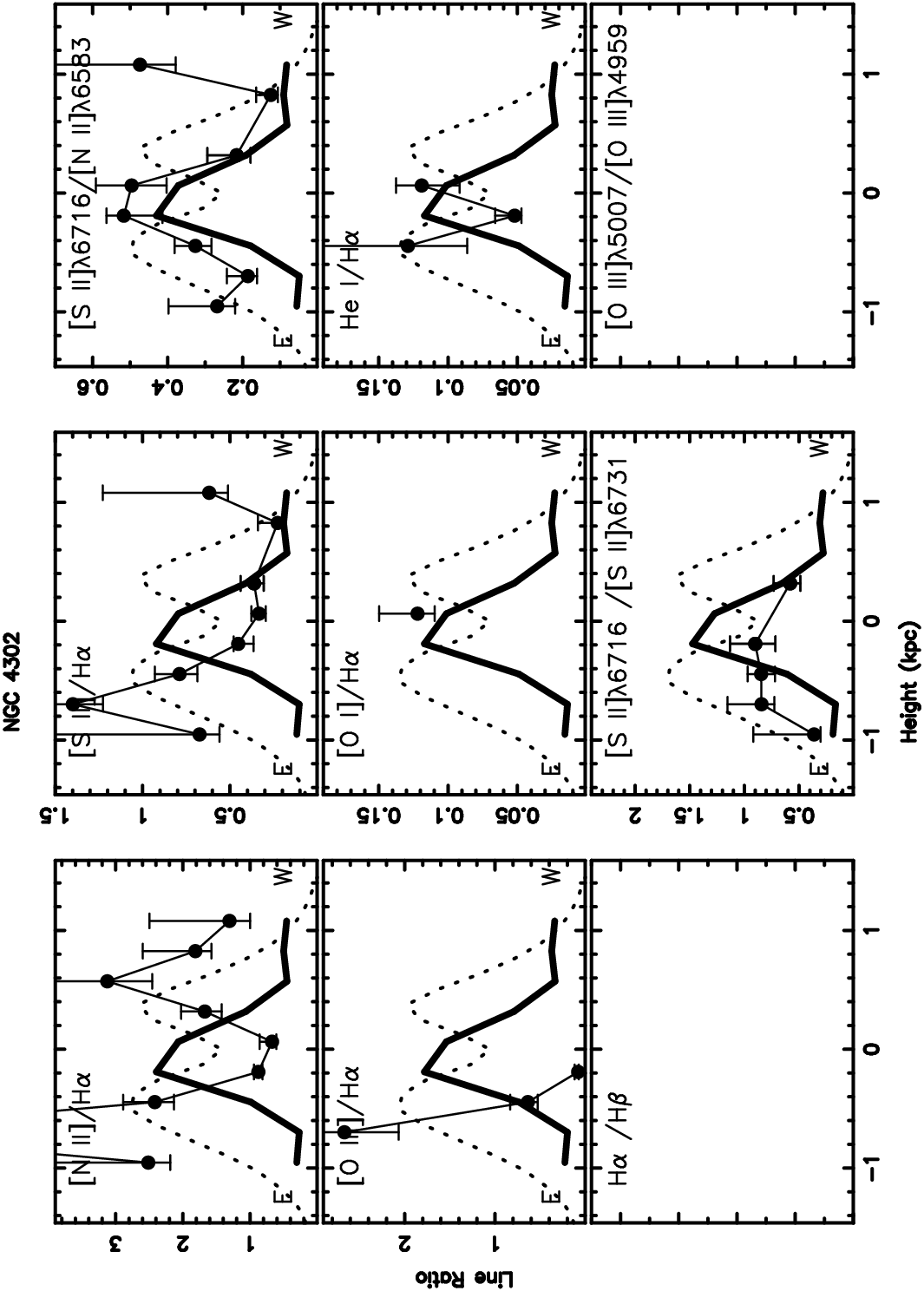}
\caption{(Cont'd.)}
\end{figure}
\epsscale{1.0}

\clearpage

\epsscale{0.65}
\begin{figure}[b]
\figurenum{3}
\plotone{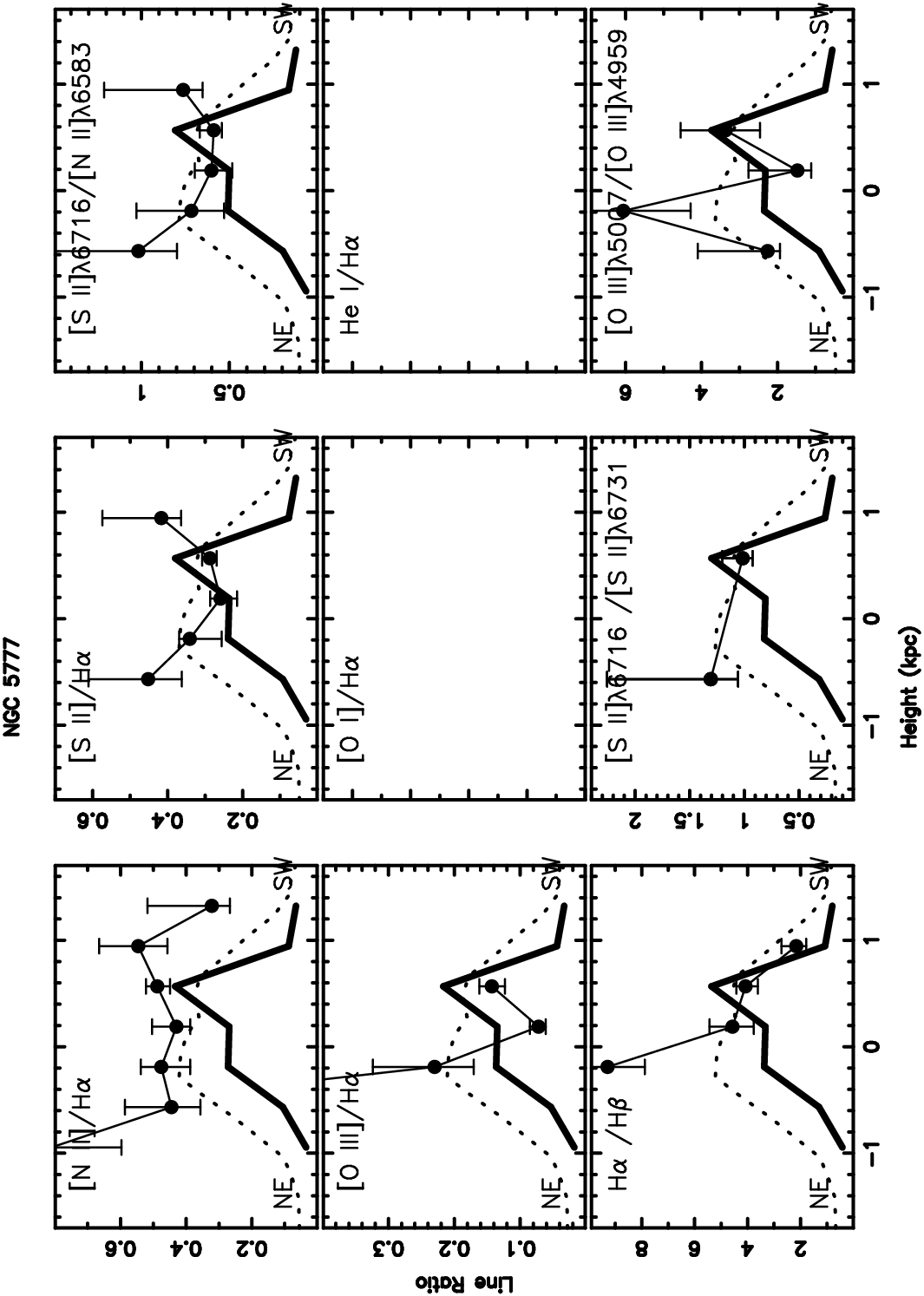}
\caption{(Cont'd.)}
\end{figure}
\epsscale{1.0}

\clearpage

%\begin{table*}[htbp]
\begin{table*}[h]
\scriptsize
\tablenum{1}
\caption{Sample}

\vskip 0.2in

\begin{tabular}{lccrccccc}
\tableline
\tableline
Galaxy &
R.A. (J2000)\tablenotemark{a} &
Dec (J2000)\tablenotemark{a} &
D$_{25}$\tablenotemark{a} &
Disk P.A.\tablenotemark{a} &
Morphological & 			
Dist.\tablenotemark{b} &                  
Incl.\tablenotemark{b} &
Reference for\\
 &
hh~mm~ss &
dd~mm~ss &
($\arcmin$) &
($^\circ$) &                               
Type\tablenotemark{a} &                               
(Mpc) &                            
($^\circ$) &
Imaging Data\\
\tableline
UGC~2092 & 02~36~30.0 & +07~18~00 &  3.16 &  32 & Scd    & 72\tablenotemark{c}   & 86\tablenotemark{d} & Pildis et al. 1994   \\
UGC~3326 & 05~39~36.0 & +77~18~00 &  3.55 &  84 & Scd    & 48\tablenotemark{c}   & 90\tablenotemark{d} & Pildis et al. 1994   \\
UGC~4278 & 08~13~59.0 & +45~44~43 &  4.68 & 172 & SB(s)d & 10.6                  & 90                  & Rand 1996            \\
NGC~2820 & 09~21~47.1 & +64~15~29 &  4.07 &  59 & SB(s)c & 20.0\tablenotemark{c} & 90                  & Paper I           \\
NGC~3628 & 11~20~16.3 & +13~35~22 & 14.79 & 104 & Sb     & 6.7                   & 87                  & Fabbiano et al. 1990 \\
NGC~4013 & 11~58~31.7 & +43~56~48 &  5.25 &  66 &  Sb    & 17                    & 90\tablenotemark{d} & Rand 1996, Paper I   \\
NGC~4217 & 12~15~50.9 & +47~05~32 &  5.25 &  50 & Sb     & 17                    & 86\tablenotemark{d} & Rand 1996            \\
NGC~4302 & 12~21~42.5 & +14~36~05 &  5.50 & 178 & Sc     & 16.8                  & 90\tablenotemark{d} & Rand 1996            \\
NGC~5777 & 14~51~18.3 & +58~58~35 &  3.09 & 144 & Sb     & 25\tablenotemark{c}   & 83\tablenotemark{d} & Pildis et al. 1994   \\
\tableline
\end{tabular}
\tablenotetext{a}{Values taken from de Vaucouleurs et al. 1991.}

\tablenotetext{b}{Values taken from Tully 1988 unless otherwise noted.}

\tablenotetext{c}{References for distances: UGC~2092, UGC~3326, \& NGC~5777 (Pildis et al.~1994), NGC~2820 (Hummel \& van der Hulst 1989).}

\tablenotetext{d}{References for inclinations: UGC~2092 \& UGC~3326 (Guthrie~1992), NGC~4013 (Bottema~1995), NGC~4217 \& NGC~4302 (Rand~1996), NGC~5777 (Guthrie 1992).}

\end{table*}

\begin{table*}[htbp]
\tablenum{2}
\caption{Observing Logs}
\vskip 0.2in
\begin{tabular}{lcccc}
\tableline
\tableline
Galaxy &
Total Exposure&
Number of &
Slit &
Distance from \\
 &
Time &
Exposures&
P.A. ($^\circ$) &
Nucleus \\
\tableline
UGC~2092 & 5.75 hours & 9  & 122 & 20\arcsec NE  \\
UGC~3326 & 4.5  hours & 7  & 107 & 16\arcsec SW  \\
UGC~4278 & 5.0  hours & 8  &  83 &  0\arcsec     \\
NGC~2820 & 5.0  hours & 7  & 139 & 19\arcsec NE  \\
NGC~3628 & 3.75 hours & 5  & 210 & 20\arcsec W   \\
NGC~4013 & 4.5  hours & 6  & 155 & 40\arcsec NE  \\
NGC~4217 & 4.5  hours & 6  & 144 & 50\arcsec SW  \\
NGC~4302 & 4.5  hours & 6  &  88 &  0\arcsec     \\
NGC~5777 & 4.5  hours & 6  &  54 & 50\arcsec SE  \\
\tableline
\end{tabular}
\end{table*}

\begin{table*}[htbp]
\tablenum{3}
\caption{Reddening Factors for Important Line Ratios if $A_V$ = 1}
\vskip 0.2in
\begin{tabular}{lc}
\tableline
\tableline
Line Ratio &
Reddening \\
\tableline
$[$N II$]~\lambda$6583/H$\alpha$  &  1.004 \\
$[$S II$]~\lambda$6716/H$\alpha$  &  1.022 \\
$[$O I$]~\lambda$6300/H$\alpha$   & 0.963 \\
He~I~$\lambda$5876/H$\alpha$   & 0.898 \\
$[$O III$]~\lambda$5007/H$\alpha$ & 0.751 \\
H$\alpha$/H$\beta$             &  1.381 \\
$[$S II$]~\lambda$6716/[N II]~$\lambda$6583   &  1.021 \\
$[$S II$]~\lambda$6716/[S II]~$\lambda$6731   & 0.998 \\
$[$O III$]~\lambda$5007/[O III]~$\lambda$4959 &  1.012 \\
\tableline
\end{tabular}
\end{table*}


\vskip 0.3in

\begin{references}
\refpar
Baldwin, J. A., Phillips, M. M., \& Terlevich, R. 1981, PASP, 93, 5
\refpar
Bland-Hawthorn, J., Freeman, K. C., Quinn, P. J., 1997, ApJ, 490, 143
\refpar
Boesgaard, A. M., \& Steigman, G. 1985, ARA\&A, 23, 319
\refpar
Bottema, R., 1995, A\&A, 295, 605
\refpar
Bregman, J. 1980, ApJ, 236, 577
\refpar
Brockelhurst, M. 1972, MNRAS, 157, 211
\refpar
Collins, J. A., Benjamin, R. A., \& Rand, R. J. 2002, ApJ, 578, 98
\refpar
Collins, J. A., \& Rand, R. J. 2001, ApJ, 551, 57
\refpar
Dahlem, M. 1997, PASP, 109, 1298
\refpar
Dahlem, M., Heckman, T. M., Fabbiano, G., Lenhert, M. D., \& Gilmore,
D. 1996, ApJ, 461, 724
\refpar
de Vaucouleurs, G., de Vaucouleurs, A., Corwin, H. G., Buta, r. J.,
Paturel, G., \& Fouqu\'e, P., 1991, Third Catalogue of Bright
Galaxies, Springer, New York
\refpar
Dettmar, R. -J. 1992, Fund. Cosmic Physics, 15, 143
\refpar
Dettmar, R. -J., \& Schultz, H. 1992, A\&A, 254, L25 
\refpar
Domg\"orgen, H., \& Dettmar, R.-J., 1997, A\&A, 322, 391
\refpar
Domg\"orgen, H., \& Mathis, J. S. 1994, ApJ, 428, 647 
\refpar
Dopita, M. A., Kewley, L. J., Heisler, C. A., \& Sutherland, R. S. 2000, ApJ, 542, 224
\refpar
Dopita, M. A., \& Sutherland, R. S., 1995, ApJ, 455, 468
\refpar
-----., 1996, ApJS, 102, 161
\refpar
Evans, I. N., \& Dopita, M. A. 1985, ApJS, 58, 125
\refpar
Fabbiano, G., Heckman, T., \& Keel, W. C., 1990, ApJ, 355, 442
\refpar
Ferguson, A., Wyse, R., \& Gallagher, J. 1996, AJ, 112, 2567
\refpar
Goad, J. W., \& Roberts, M. S. 1981, ApJ, 250, 79
\refpar
Golla, G., Dettmar, R.-J., \& Domg\"orgen, H., 1996, A\&A, 313, 439
\refpar
Guthrie, B. N. G., 1992, A\&AS, 93, 255
\refpar
Haffner, L. M., Reynolds, R. J., \& Tufte, S. L. 1999, ApJ, 523, 223
\refpar
Hartquist, T. W., \& Morfill, G. E. 1986, ApJ, 311, 518
\refpar
Houck, J. C., \& Bregman, J. N. 1990, ApJ, 352, 506
\refpar
Hummel, E., \& van der Hulst, J. M. 1989, A\&AS 81,51 
\refpar
Kaler, J. B. 1976, ApJS, 31, 517
\refpar
Keppel, J. W., Dettmar, R.-J., Gallagher, J. S., \& Roberts, M. S.
1991, ApJ, 374, 507
\refpar
Lerche, I., \& Schlickeiser, R. 1982 A\&A, 107, 148.
\refpar
Martin, P. G. 1988, ApJS, 66, 125
\refpar
Mathis, J. S. 1986, ApJ, 301, 423
\refpar
-----. 2000, ApJ, 544, 347
\refpar
McCall, M. L., Rybski, P. M., \& Shields, G. A. 1985, ApJS, 57, 1
\refpar
Meyer, D. M., Cardelli, J. A., \& Sofia, U. J. 1997, ApJ, 490, L103
\refpar
Miller, S. T., \& Veilleux, S. 2003a, ApJS, submitted (Paper I)
\refpar
Osterbrock, D., 1989, Astrophysics of Gaseous Nebulae and Active
Galactic Nuclei (Mill Valley: University Science)
\refpar
Osterbrock, Tran, \& Veilleux 1992, ApJ, 389, 196
\refpar
Otte, B., \& Dettmar, R.-J., 1999, A\&A, 343, 705
\refpar
Otte, B., Gallagher, J. S., \& Reynolds, R. J. 2002, ApJ, 572, 823
\refpar
Otte, B., Reynolds, R. J., Gallagher, J. S., \& Ferguson, A. M. N. 2001, ApJ, 560, 207
\refpar
Parker, E. N. 1992, ApJ, 401, 137.
\refpar
Pildis, R. A., Bregman, J. N., \& Schombert, J. M. 1994a, ApJ 423, 190
\refpar
-----. 1994b, ApJ 427, 160 (PBS)
\refpar
Rand, R. J. 1996, ApJ, 462, 712 (R96)
\refpar
-----. 1997, ApJ, 474, 129
\refpar
-----. 1998, ApJ, 501, 137
\refpar
-----. 2000, ApJ, 537, 13
\refpar
Rand, R. J., Kulkarni, S. R., \& Hester, J. J. 1990, ApJ, 352,L1
(erratum 362, L35)
\refpar
Raymond, J. C. 1992, ApJ, 384, 502
\refpar
Reynolds, R. J. 1984, ApJ, 282, 191
\refpar
-----. 1985a, ApJ, 294, 256
\refpar
-----. 1985b, ApJ, 298, L27
\refpar
------. 1992, ApJ, 392, L35
\refpar
Reynolds, R. J., Haffner, L. M., \& Tufte, S. L. 1999, ApJ, 525, L21
\refpar
Reynolds, R. J., Sterling, N. C., Haffner, L. M., \& Tufte,
S. L. 2001, ApJ, 548, 221
\refpar
Reynolds, R. J., \& Tufte, S. L. 1995, ApJ, 439, L17
\refpar
Shull, J. M., \& McKee, C. F., 1979, ApJ, 227, 131
\refpar
Slavin, J., McKee, C., \& Hollenbach, D., 2000, ApJ, 541, 218
\refpar
Slavin, J. D., Shull, J. M., \& Begelman, M. C. 1993, ApJ, 407, 83 
\refpar
Sokolowski 1994, PhD thesis, Rice University 
\refpar
Stasinski, G. 1982, A\&AS, 48, 299
\refpar
T\"ullman, R. \& Dettmar, R.-J. 2000, A\&A, 362, 119
\refpar
T\"ullman, R., Dettmar, R.-J., Soida, M., Urbanik, M., \& Rossa,
J. 2000, A\&A, 364, L36
\refpar
Tully, R. B., 1988, Nearby Galaxies Catalog (Cambridge: Cambridge
University Press)
\refpar
Veilleux, S. 2001, in IAU Coll. 184, AGN Surveys, eds. R.F. Green, E.Ye. Khachikian, and D.B. Sanders, ASP Conf. Series., in press (astro-ph/0201118)
\refpar
Veilleux, S., Cecil, G., \& Bland-Hawthorn, J. 1995, ApJ, 445, 152
\refpar
Veilleux, S., \& Osterbrock, D. E. 1987, ApJS, 63, 295
\end{references}
\end{document}